\documentclass[aps,prl,twocolumn]{revtex4-1}
\usepackage{graphicx}
\usepackage{amssymb}
\usepackage{color}
\usepackage{float}

\begin{document}

\title{Optical Probes of the Quantum-Entangled Triplet-Triplet State in a Heteroacene Dimer}
\author{Souratosh Khan}
\affiliation{Department of Physics, University of Arizona
Tucson, AZ 85721}
\author{Sumit Mazumdar}
\affiliation{Department of Physics, University of Arizona}
\affiliation{Department of Chemistry and Biochemistry, University of Arizona}
\affiliation{College of Optical Sciences, University of Arizona}
\date{\today}

\begin{abstract}
\noindent  
The nature and extent of the spin-entanglement in the triplet-triplet biexciton with total spin zero in correlated-electron 
$\pi$-conjugated systems continues to be
an enigma. Differences in the ultrafast transient absorption spectra of free triplets versus the triplet-triplet 
can give a measure of the entanglement. This, however, requires theoretical understandings of transient absorptions
from the optical spin-singlet, the lowest spin-triplet exciton as well as from the triplet-triplet state, 
whose spectra are often overlapping and hence difficult to distinguish. We present a many-electron theory of the electronic structure
of the triplet-triplet, and of 
complete wavelength-dependent excited state absorptions (ESAs) from all three states in a heteroacene dimer of interest
in the field of intramolecular singlet fission. The theory allows direct comparisons of ESAs with existing experiments as well as experimental
predictions, and gives physical understandings of transient absorptions within a pictorial exciton basis that 
can be carried over to other experimental systems.
\end{abstract}
\maketitle

\section{Introduction}
Carbon-based
$\pi$-conjugated systems have been the testing ground for quantum chemical many-body approaches since the beginning of quantum
chemistry \cite{Salem66a}.
The detection of an even parity, dipole forbidden 2$^1$A$_g^-$
state below the lowest optical 1$^1$B$_u^+$ exciton in linear polyenes led to a paradigm shift in our understanding of $\pi$-conjugated
systems, providing a clear demonstration of the dominant role of Coulomb repulsion on their electronic structures \cite{Kohler73a,Hudson82a}.
As has been explicitly shown within correlated $\pi$-electron theory \cite{Schulten76a,Ramasesha84b,Ramasesha84c},
the 2$^1$A$_g^-$ and other low lying even parity states in polyenes are covalent bound states of two spin triplet excitons T$_1$, 
hereafter referred to as the triple-triplet biexciton $^1$(TT)$_1$,
whose spin angular momenta  are quantum-entangled to yield a spin singlet. 
More recently, low lying triplet-triplet states have been theoretically predicted in large polycyclic hydrocarbons \cite{Aryanpour14a}
and graphene nanoribbons \cite{Goli16a}. 

Similar $^1$(TT)$_1$ state has acquired considerable importance as the dominant intermediate in the photophysical process of
singlet fission, hereafter SF, in which 
the optically generated spin-singlet exciton S$_1$ dissociates into two lowest triplet excitons T$_1$ in two or more steps \cite{Smith13a}. 
The process is being
intensely investigated, because of its potential utilization as a means to double the photoconductivity in organic solar cells.
The overall SF process is usually written as S$_0$ + S$_1 \to ^1$(TT)$_1 \to$ T$_1$ + T$_1$, where S$_0$ 
refers to the ground state.

Experimental confirmation of SF is usually done from transient absorption (TA) spectroscopy: paired ultrafast decay of the TA from S$_1$ with the 
concomitant appearance of TA from T$_1$ would be the signature of SF. Reevaluations of the interpretations of
longstanding experimental observations are currently in progress \cite{Yong17a,Stern17a,Weiss16a,Tayebjee17a,Basel17a}, because of realizations that
(i) the $^1$(TT)$_1$ may be more longlived than believed until now, and (ii) spectroscopic signatures previously assigned to T$_1$ may actually
originate from the $^1$(TT)$_1$. Precise identification of T$_1$ versus $^1$(TT)$_1$ from TA spectroscopy is therefore crucial for determining
whether SF has been complete. Simultaneously, the difference in the TA spectra of T$_1$ and $^1$(TT)$_1$  
is a measure of the spin-entanglement in the latter, and theoretical and experimental knowledge of the extent of this entanglement 
 can have 
practical applications in widely varying research fronts such as quantum information theory, organic spintronics, and phosphorescent 
light emitting diodes.
%

In the present paper we develop a broad theory of the quantum-entangled electronic structure of the $^1$(TT)$_1$ in heteroacene dimers
of TIPS-pentacene (TIPS-P) and TIPS-tetracene (TIPS-T), PTn, linked by n = 0, 1 and 2 phenyl groups, respectively (see Fig.~1).
We present computational results of ESAs from S$_1$, T$_1$ and $^1$(TT)$_1$ that allow direct comparisons with 
existing experimental results \cite{Sanders16c}, as well as making experimental predictions. Most importantly, our theoretical approach gives physically
intuitive undertanding of all eigenstates and ESAs within a pictorial exciton basis introduced previously \cite{Chandross99a,Khan17b, Khan17c}.
The lack of inversion symmetry in PTn makes the present study more general than our previous study of similar dimers of TIPS-P, BPn. 
Consequently, the physical interpretations of eigenstates and ESAs developed here can be carried over to other molecular systems of interest. Finally, the
smaller sizes of PTn relative to BPn allow investigations of upto n = 2 which was not possible for BPn. 
We will see that with increasing n there occurs a gradual decrease in entanglement.
It is important to recall in this context that spin quintet (as opposed to spin singlet) triplet-triplet states
have been observed for n $>$ 2 recently \cite{Tayebjee17a}.

\begin{figure}
\includegraphics[width=3.in]{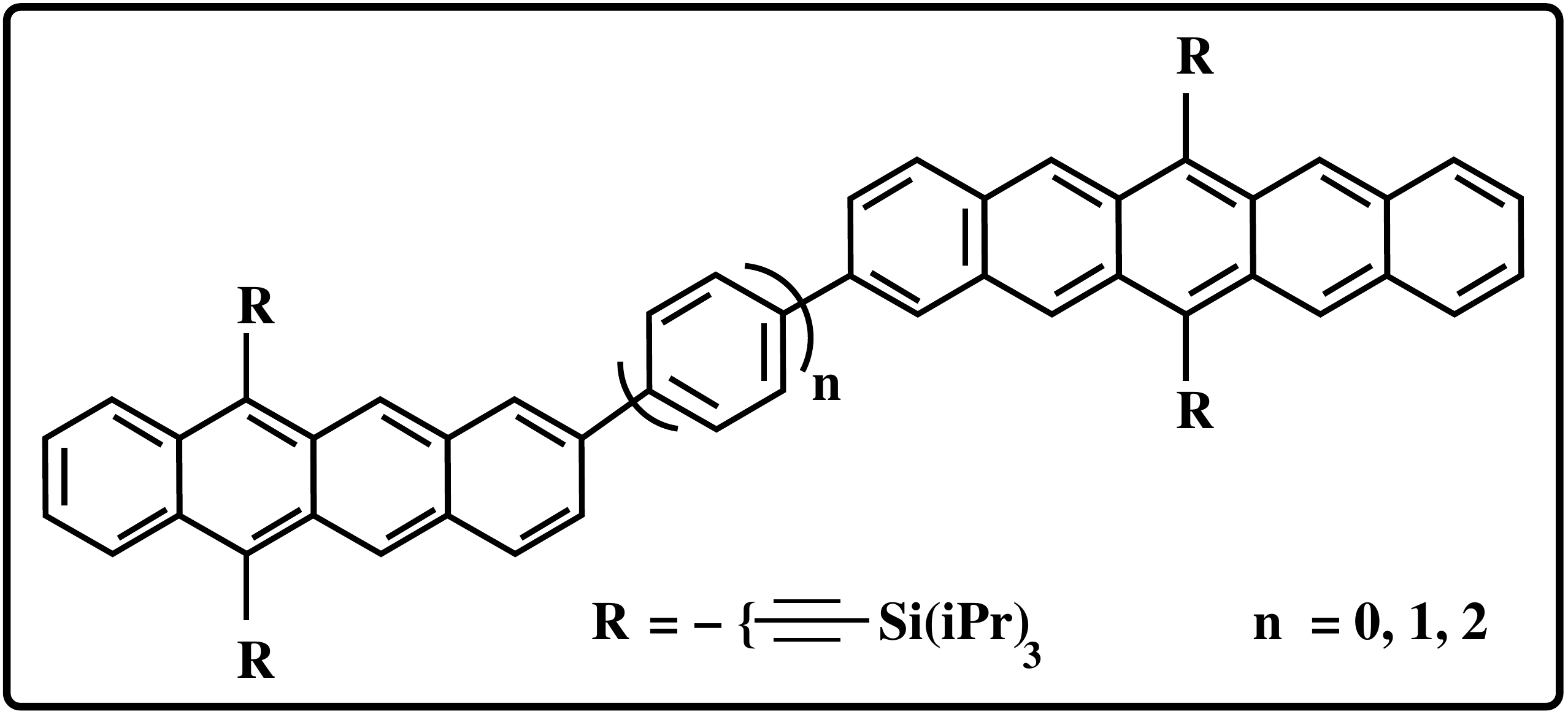}
\caption{PTn dimer: TIPS-pentacene and TIPS-tetracene molecules bridged with n = 0, 1 and 2 phenyl spacers.}
\label{PTn}
\end{figure}

\section{Theoretical model, parametrization and approach}
Our calculations are within the $\pi$-electron Pariser-Parr-Pople (PPP) Hamiltonian \cite{Pariser53a,Pople53a},
\begin{equation}
\label{PPP_Ham}
H_{\mathrm{PPP}}=H_{\mathrm{intra}}+H_{\mathrm{inter}}\,,
\end{equation}
\begin{eqnarray}
\label{intra_Ham}
 H_{\mathrm{intra}}=\sum_{\mu,\langle ij \rangle,\sigma}t_{ij}^{\mu}(\hat{c}_{\mu i\sigma}^{\dagger}\hat{c}_{\mu j\sigma}^{}+\hat{c}_{\mu j\sigma}^\dagger \hat{c}_{\mu i\sigma}^{}) \\
\nonumber
+ U\sum_{\mu, i}\hat{n}_{\mu i\uparrow} \hat{n}_{\mu i\downarrow} + \sum_{\mu,i<j} V_{ij} (\hat{n}_{\mu i}-1)(\hat{n}_{\mu j}-1) \\
\nonumber
\end{eqnarray}
\begin{eqnarray}
\label{inter_Ham}
 H_{\mathrm{inter}}=\sum_{\mu\neq\mu',ij,\sigma}t^{inter}_{ij}
(\hat{c}_{\mu i\sigma}^\dagger \hat{c}_{\mu' j\sigma}^{}+\hat{c}_{\mu' j\sigma}^\dagger \hat{c}_{\mu i\sigma}^{}) \\
\nonumber 
+ \frac{1}{2}\sum_{\mu\neq\mu',ij} V_{ij}^{inter}(\hat{n}_{\mu i}-1)(\hat{n}_{\mu' j}-1)\,,\hspace{0.5in}
\end{eqnarray}
Here $H_{\mathrm{intra}}$ describes interactions within the individual TIPS-P and TIPS-T monomers and phenyl linkers, while
$H_{\mathrm{inter}}$ describes the interactions between these molecular units.
Our approach allows
descriptions of all many-body eigenstates in terms of a physical, pictorial {\it exciton basis} \cite{Chandross99a,Khan17c,Khan17b}. In the above 
$\hat{c}^{\dagger}_{\mu i\sigma}$ creates a $\pi$-electron of spin $\sigma$ on carbon (C) atom $i$ within the monomer unit $\mu$,  
$\hat{n}_{\mu i\sigma} = \hat{c}^{\dagger}_{\mu i\sigma}\hat{c}_{\mu i\sigma}^{}$ is the number of electrons of spin $\sigma$,
and $\hat{n}_{\mu i}=\sum_{\sigma} \hat{n}_{\mu i\sigma}$. 
The intraunit nearest neighbor hoppings $t_{ij}^{\mu}$ are taken to be
$-2.4$ eV and $-2.2$ eV for the peripheral and internal carbon bonds of the TIPS-P and TIPS-T units, respectively, 
based on (i) first principle calculations \cite{Houk01a} that determined the corresponding average bond lengths to be
1.40 $\mathring{\textrm{A}}$ and  1.46 $\mathring{\textrm{A}}$, respectively, and (ii) a widely accepted bond length-hopping integral relationship 
\cite{Ducasse82a}.
The C-C hopping integrals corresponding to the internal bonds in the phenyl ring and to the triple bond in the 
TIPS group are taken to be $-2.4$ eV and $-3.0$ eV, respectively \cite{Ducasse82a}.
The inter-unit hopping integral $t^{inter}_{ij}$ is fixed at $-2.2$ eV for the bulk of our calculations, which assumes a planar geometry. 
Rotational twists between units can be taken care of by reducing $t^{inter}_{ij}$, as is discussed later.
$U$ and $V_{ij}$ are the on-site and long range Coulomb repulsions.
We employ the modified Ohno parameterization for the latter, $V_{ij} = U/\kappa\sqrt{1+0.6117 R_{ij}^2}$, where $\kappa$ is an effective dielectric constant \cite{Chandross97a}. 
Based on previous work \cite{Khan17b,Khan17c},
we calculate absorption spectra in the spin singlet subspaces, ground and excited, with  $U=6.7$ eV and $\kappa=1.0$, while all triplet and triplet-triplet excited state
absorption spectra are calculated with $U=7.7$ eV and $\kappa=1.3$ \cite{sm}.

Our PPP calculations are electron-only and ignores relaxations of excited state energies 
due to electron-vibration coupling. 
The calculations of ESAs from the correlated-electron eigenstates of the PPP Hamiltonian require solving configuration interaction Hamiltonian matrices 
that have dimensions several times 10$^6$ (see below). 
Including nuclear relaxations in calculations of ESAs to states that are at twice the energy of the singlet exciton, 
or that are from the highly correlated $^1$(TT)$_1$, which has contributions from single to quadruple many-electron excitations (see below), 
is currently outside the scope of correlated-electron calculations. 
Thus completely quantitative fittings of calculated and experimental ESA energies are not to be expected. 
Because of the strong Coulomb interactions that localize excitations, 
we expect the errors in the calculated ESA energies to be small enough to achieve our major goals, 
viz., to determine the differences between the ESAs (i) from the the optical singlet, 
free triplet and the $^1$(TT)$_1$ on the one hand, and (ii) from PTn versus BPn, on the other, at a {\it qualitative} level.

\begin{figure*}[t]
\includegraphics[width=7.5in]{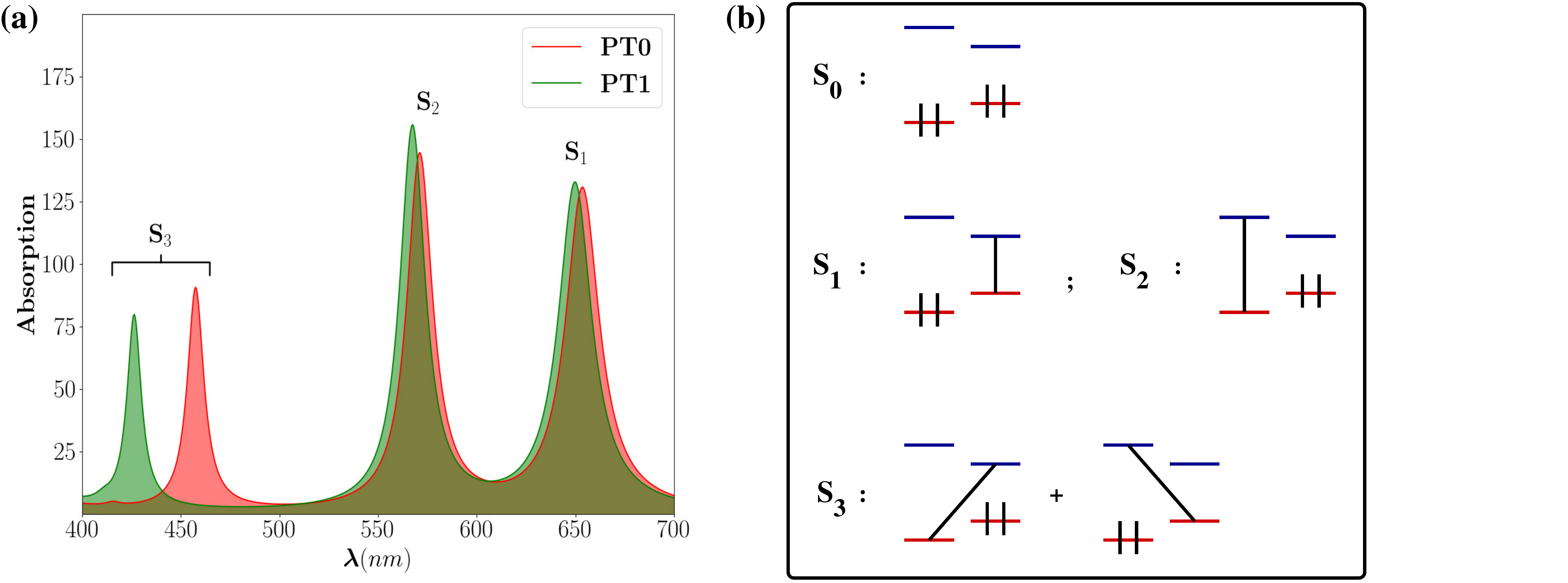}
\caption{
(a) Calculated electronic optical absorption spectra of PT0 (red), PT1 (green) with $U=6.7$ eV, $\kappa=1.0$.
(b) Most dominant exciton basis configurations in S$_0$, S$_1$, S$_2$ and S$_3$ in PT0 and PT1. 
Only the HOMO (red) and LUMO (blue), and their occupancies by electrons are shown. The unit with the smaller (larger)
HOMO-LUMO gap is TIPS-P (TIPS-T). The black lines connecting bonding and antibonding MOs are spin singlet excitations.}
\label{grst}
\end{figure*}

We use the multiple reference singles and doubles configuration interaction (MRSDCI) approach \cite{Tavan87a} 
to obtain all correlated state energies and wavefunctions. 
Our basis functions are obtained by solving the PPP hamiltonian (Eq 2) within the Hartree-Fock (HF) approximation in the limit $H_{inter}=0$ and 
occupying the HF MOs with single, double etc excitations from the HF ground state. 
Computational limitations prevent us from including the entire active space of 48 MOs in PT0 and 54 MOs in PT1. 
We retain 24 MOs (12 bonding and 12 antibonding), including the two highest (lowest) bonding (antibonding) phenyl MOs.
For each eigenstate, we perform an initial double-CI calculation in the complete space of double excitations, now for $H_{inter} \neq 0$.
We then discard the singly and doubly excited configurations whose contributions to the double-CI wavefunction are below a cutoff value. We
retain the dominant $N_{ref}$ singly and doubly excited configurations, and perform a CI calculation in which double excitations 
from these $N_{ref}$ configurations are included, thereby effectively including the most dominant triple and quadruple excitations. 
These triple and quadruple excitations, in turn, have large CI matrix elements with new single and double excitations that had not been
included in the original $N_{ref}$ reference configurations. The new single and double excitations are now included in the updated $N_{ref}$ and the
procedure is repeated iteratively until convergence is reached.
In the Supplemental Material \cite{sm}, we have given the convergence criterion and the final $N_{ref}$ and $N_{total}$, the overall dimension of the Hamiltonian matrix for relevant eigenstates of all molecules we have investigated.
In all cases our CI matrices are several times 10$^6$ in size.

\section{Results and Discussion}
\subsection{Ground state absorption}

The calculated ground state absorption spectra of PT0 and PT1, shown in Fig.~\ref{grst}(a), agree very closely with the experimental absorption spectra (see Fig.~2 in reference \onlinecite{Sanders16c};
vibrational sidebands seen experimentally are not expected from computations based on the purely electronic PPP model).
The absorptions labeled S$_1$ and S$_2$ match very closely with the spectra in monomer solutions of TIPS-pentacene \cite{Platt09a,Ramanan12a}  and TIPS-tetracene \cite{Stern15a} (absorption maxima at 660 nm and 566 nm, respectively).
The absorption bands S$_3$ are absent in the monomers but have been seen experimentally in the dimers \cite{Sanders16c}. In Fig.~\ref{grst}(b) we have shown the dominant exciton basis contributions
to the ground state S$_0$, and the excited states S$_1$, S$_2$ and S$_3$, respectively, for both PT0 and PT1. We have included only the highest occupied and lowest unoccupied (HOMO
and LUMO) MOs in our depiction of the exciton basis configurations, as contributions by excitations from lower bonding MOs or to higher
antibonding MOs are weak for these lowest singlet excited states. As indicated in Fig.~\ref{grst}(b), the final states S$_1$ and S$_2$ of the 
two lowest energy absorptions are to two {\it distinct}
eigenstates with Frenkel exciton (FE) character localized 
on the individual TIPS-P and TIPS-T monomers, and {\it not to a single delocalized eigenstate that is a linear superposition of the two, as had been claimed before} \cite{Sanders16c} 
(see reference \onlinecite{sm} for complete wavefunctions).
This is different from the symmetric dimers BPn, where the lowest absorption is indeed a superposition state of degenerate FEs on identical monomers. 
As also shown in Fig.~\ref{grst}(b), the final state S$_3$ of the highest energy absorption is a charge-transfer (CT) state with 
charge transfer in both directions with equal probability. A similar CT absorption (labeled S$_2$ there), is also found in BPn \cite{Khan17b}.
Distinct FE and CT excitations, as opposed to a strong superposition, are a sign of strong electron correlations \cite{Chandross99a}. 
Absorptions from FE excitations shown in Fig.~\ref{grst}(a) are polarized along the short axes of the monomers \cite{Sony07a,Khan17b} while CT excitations are predominantly polarized along the long molecular axis of the dimer. 

While TIPS-T and PTn do not have inversion symmetry, they possess charge-conjugation symmetry in the limit of nearest neighbor-only electron hopping, and transition dipole matrix elements
are nonzero only between initial and final states with opposite charge-conjugation symmetries. 
Additionally, for strong Coulomb correlations,
the lowest eigenstates that are optically allowed from the ground state are necessarily ionic in the language of valence bond theory.
Conversely, excited states that are predominantly covalent are one-photon forbidden. Our calculations using the exciton basis do not
use any symmetry. We have found, however, that there is no mixing between one-photon states optically allowed from the
ground state and two-photon states that are reached in excited state absorption.

\begin{figure}[t]
\includegraphics[width=3.5in]{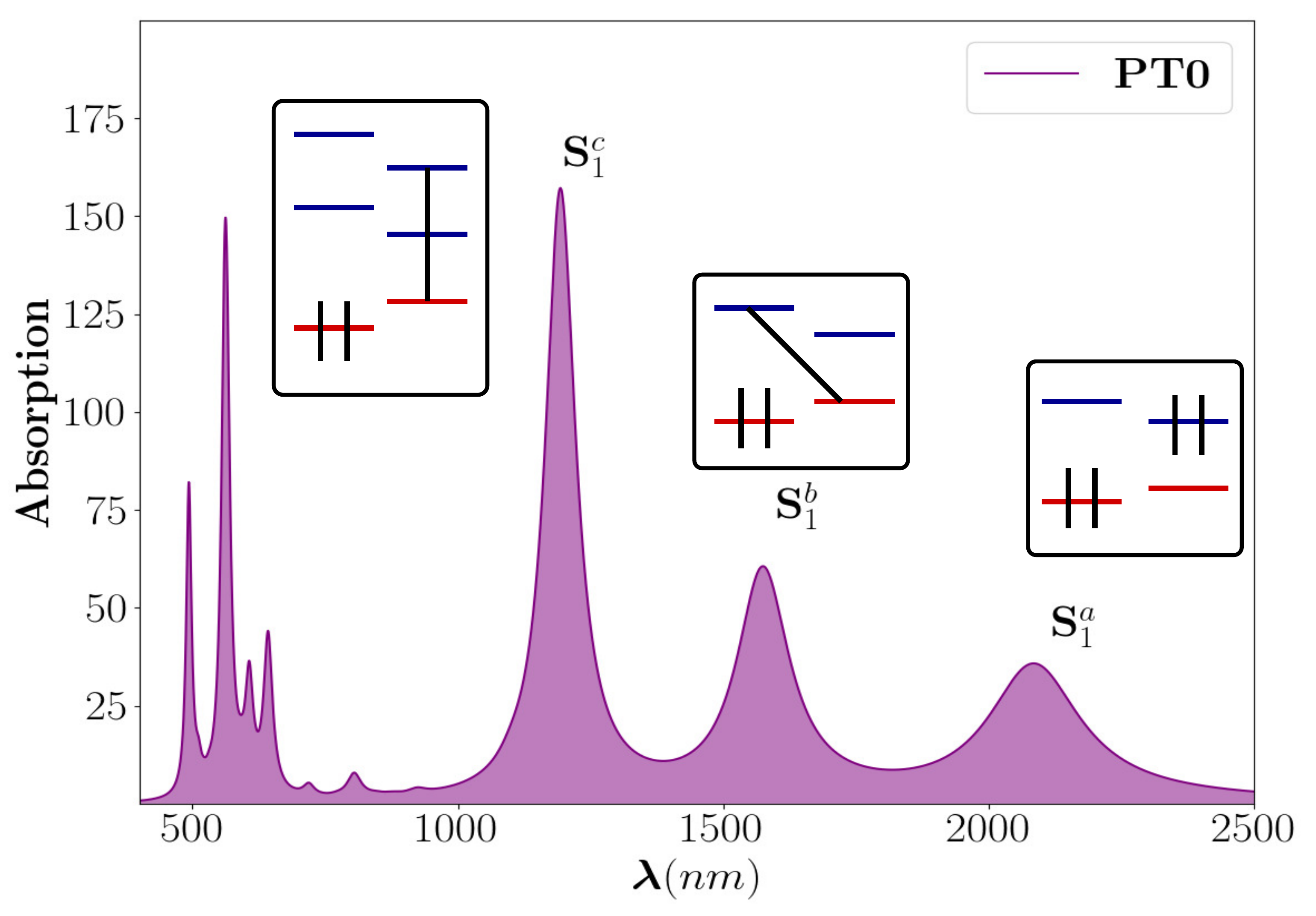}
\caption{Calculated singlet ESA from S$_1$ in PT0 with $U=6.7$ eV and $\kappa=1.0$. 
Insets show the dominant contributions to the final states of the absorptions (see text).  
S$^a_1$ is an intramonomer 2e-2h covalent state that is analogous to the 2$^1$A$_g^-$ state in linear polyenes.}
\label{singletESA}
\end{figure}

\subsection{Transient Absorptions}

We have calculated all ESAs relevant for understanding existing \cite{Sanders16c} and future transient absorption measurements in this 
heteroacene dimer. The advantage of the exciton basis representation is that
ESAs of weakly coupled units can be understood as intraunit and interunit transitions, where the intraunit excitations can be further classified as
one electron-one hole and two electron-two hole 
(1e-1h and 2e-2h), respectively \cite{Psiachos09a}.
Further, it also allows predictions of the polarizations of ESAs, based on the MOs involved in the dominant excitations. These predictions
are then confirmed from explicit computations, leading to additional one-to-one correspondence between the calculated ESAs and the wavefunction
analyses. In general, 
intraunit HOMO $\to$ LUMO (LUMO $\to$ LUMO+1) transitions are polarized along the short (long) axes of the dimer molecule, 
while all interunit excitations are naturally polarized along the long molecular axis.   
In the following sections, we discuss calculated ESAs from S$_1$, T$_1$ and $^1$(TT)$_1$.
\begin{figure*}[t]
\includegraphics[width=7.5in]{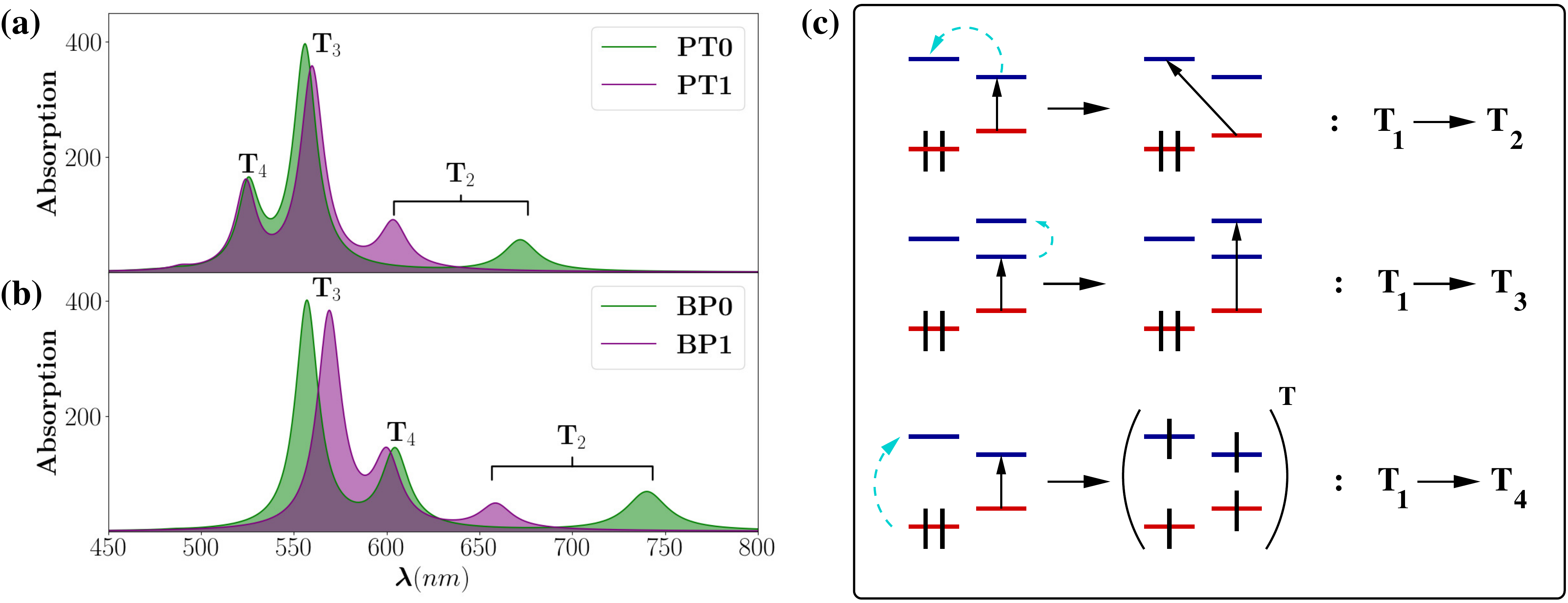}
\caption{Calculated triplet ESA spectra of (a) PT0 (green) and PT1 (purple); and (b) BP0 (green) and BP1 (purple) for $U=7.7$ eV and $\kappa=1.3$. 
(c) Dominant exciton basis contributions to the final states T$_2$, T$_3$ and T$_4$ in PTn. See reference \onlinecite{sm} for the descriptions of the
complete wavefunctions.
The arrow connecting MOs are triplet excitations.
}
\label{triplet}
\end{figure*}
\vskip 0.5pc
\noindent \underbar {\it (a) Singlet (S$_1$) ESA}. In Fig.~\ref{singletESA}, we have shown calculated ESA spectrum from the singlet optical exciton S$_1$
for PT0. The calculated ESA spectrum of PT1 is largely similar,
with only slightly shifted wavelengths \cite{sm}.
We find singlet ESAs in
the short-wave infrared (SWIR) to final states S$^a_1$ and S$^b_1$, in the near IR (NIR) to final state S$^c_1$, and in the visible. Absorptions
in the visible are to many different intramonomer excitations, and are strongly overlapping with ESAs from T$_1$ and $^1$(TT)$_1$ (see below).
Experimentally, in PTn, only the TAs in the visible from S$_1$ have been detected until now \cite{Sanders16c}; extending the experiments
to longer wavelengths will make distinguishing between S$_1$ and $^1$(TT)$_1$ simpler as in BPn \cite{Trinh17a} and aggregates \cite{Pensack16a,Walker13a} or crystalline films \cite{Grieco18a} of TIPS-P.   
Fig.~\ref{singletESA} also gives the dominant contributions to the final states of these ESAs. We discuss the ESAs in increasing order of energy.   

(i) S$^a_1$ is predominantly ($\sim$ 60$\%$) 2e-2h, (HOMO$\rightarrow$LUMO)$_P$ $\otimes$ (HOMO$\rightarrow$LUMO)$_P$ double excitations within the TIPS-P unit. 
This state is the pentacene monomer triplet-triplet excitation that corresponds to the 2$^1$A$_g^-$ of polyenes \cite{Khan17b}. 
Within valence bond theory, it is a covalent eigenstate \cite{Schulten76a,Ramasesha84b,Ramasesha84c} whose low energy is a consequence of strong Coulomb correlations.
ESA to S$^a_1$
is polarized predominantly along the short axis of the molecule. 

(ii) The S$_1$ $\rightarrow$ S$_{1}^{b}$ ESA is primarily ($\sim 50$\%) CT in character and is strongly polarized along the long axis of the molecule. Not surprisingly, the final state is energetically
proximate to S$_3$ in Fig.~\ref{grst}, albeit of opposite parity.

(iii) Finally, the NIR absorption to S$^c_1$ is once again intraunit, LUMO (HOMO-1) $\rightarrow$ LUMO+1 (HOMO) excitation within the TIPS-P monomer.
This absorption is polarized along the long axis of the molecule. The corresponding absorption in BP0 \cite{Khan17b} has been observed experimentally \cite{Trinh17a}.
Given that the singlet exciton S$_1$ is localized on the TIPS-P monomer, it should not be surprising that our calculated
singlet ESA spectrum is very similar to that calculated previously for BPn \cite{Khan17b}. There is, however, a strong difference in the wavefunctions 
of the final states. While in BPn the two lowest energy ESAs are to strong superpositions of the lowest 2e-2h and CT configurations, in the present
asymmetric dimer the mixing between these two clases of configurations is very weak. 
With the addition of a phenyl linker, the ESAs in PT1 are blue-shifted with an increased energy difference between S$^a_1$ and S$^b_1$.  
A complete description of the final states S$^a_1$, S$^b_1$ and S$^c_1$ and the ESA spectrum of PT1 can be found in
Supplemetal Material \cite{sm}.

ESAs S$_1$ $\to$ S$_1^c$ and  S$_1$ $\to$ S$_1^a$ are intraunit within the TIPS-P component of PTn and they should therefore 
occur also in the TIPS-P monomer. 
We have verified this from our ESA calculations of the TIPS-P monomer (see Fig.~S2 in Supplemental Material \cite{sm}). 
It is only recently this long wavelength region has been probed in monomer studies. 
Singlet ESAs in solution studies of TIPS-P
have been found close  to 1200 nm (corresponding to
S$_1$ $\to$ S$_1^c$) \cite{Walker13a}, and even more recently in the mid-IR region \cite{Grieco18a}. 
The latter has been attributed to the transition to a doubly excited state \cite{Grieco18a}, in agreement with our assignment in Fig.~\ref{singletESA}.

\begin{figure*}[t]
\includegraphics[width=7.5in]{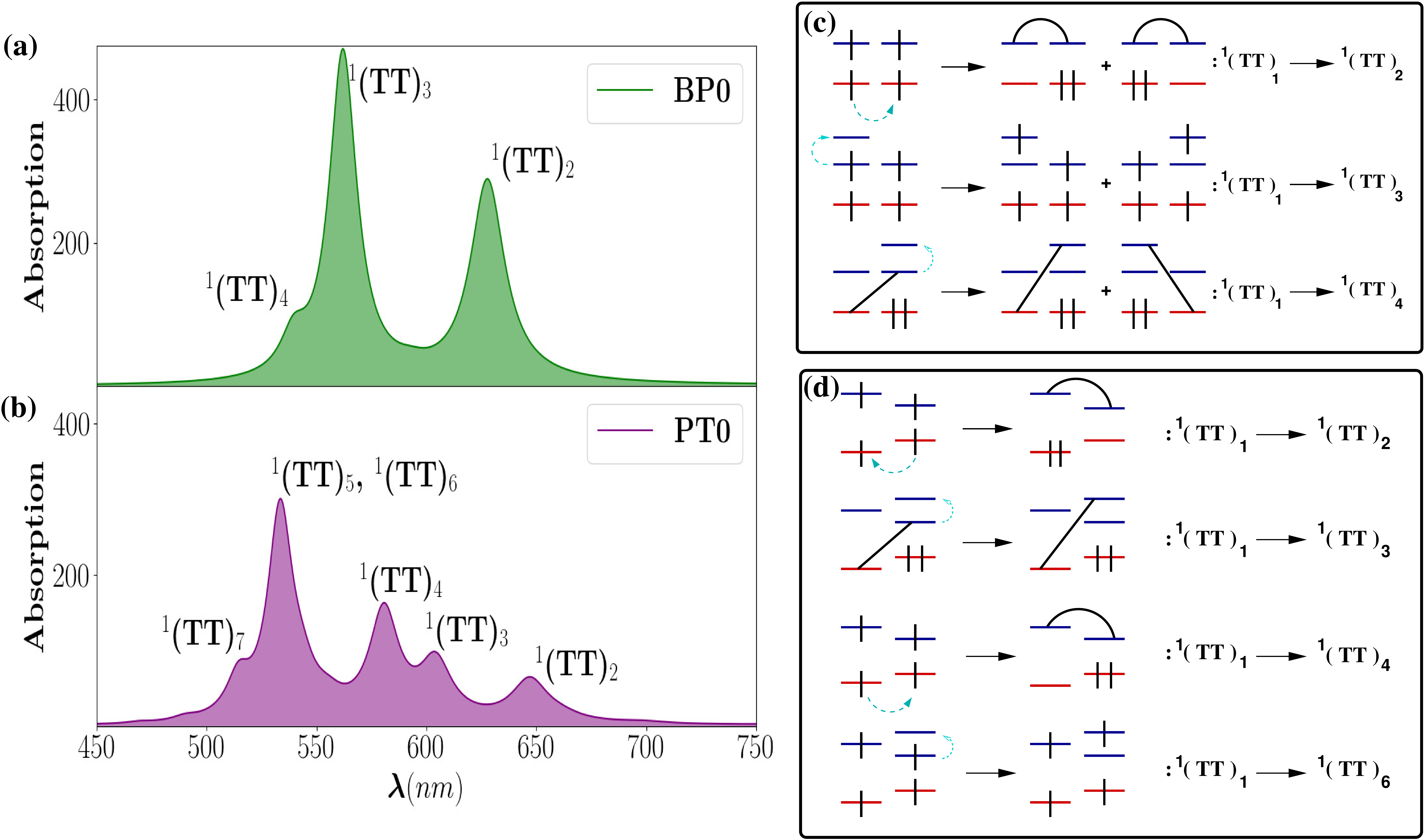}
\caption{ 
Calculated triplet-triplet ESA spectrum of (a) BP0 and (b) PT0 for $U=7.7$ eV and $\kappa=1.3$.
(b) Dominant configurations to the final states in ESA spectra of BP0. The blue broken arrows indicate the nature of 
the transition from the ``TT'' or CT component of the $^1$(TT)$_1$ state.
(d) Transitions in the triplet-triplet manifold that yield non-degenerate configurations in the final states of the ESA spectrum in PT0.
The black lines connecting MOs are again spin-singlet bonds.
}
\label{TTPA}
\end{figure*} 
\vskip 0.5pc
\noindent \underbar {\it (b) Triplet ESA.}
Like S$_1$, T$_1$ is also primarily localized on the TIPS-P unit. We find it at 0.88 eV in PTn, which is to be compared against the
calculated T$_1$ energy of 0.89 eV in the TIPS-P monomer \cite{sm}.
Its counterpart with the excitation on the TIPS-T unit is $\sim$ 0.3 eV higher in energy.
Sanders {\it et al.} have determined triplet populations in both the constituent units even at long timescales
($\sim$ 100 $\mu$s) with no triplet exciton transfer from  TIPS-T to TIPS-P \cite{Sanders16c}.
Experimentally, the absorption cross-section from the triplet in pentacene is much larger than that from tetracene
\cite{Stern15a, Walker13a}. This is confirmed in our calculations, as shown in Fig.~S7 of reference \onlinecite{sm}, where we have superimposed
triplet ESAs from TIPS-P and TIPS-T monomers. The triplet absorptions from the two molecules are largely overlapping at visible in wavelength,
but those from TIPS-T are significantly weaker. Based on the overlapping wavelengths and the much weaker strengths of the absorptions from
TIPS-T, we conclude that experimental triplet ESAs from PTn will be dominated by T$_1$. 
The calculated ESA spectra from T$_1$ in PT0 and PT1 are shown in  Fig.~\ref{triplet}(a).
The calculated triplet ESA spectra for BP0 and BP1 are shown in Fig.~\ref{triplet}(b) for comparison.
Fig.~\ref{triplet}(c)  gives the dominant contributions to the final states of Fig.~\ref{triplet}(a).
We identify three distinct absorptions in the triplet manifold.

(i) The lowest energy absorptions from T$_1$ is to CT states T$_2$, in all four cases, PT0 and PT1, BP0 and BP1. Not surprisingly, the absorptions
occur at longer wavelengths (lower energies) in the bipentacenes.

(ii) Following this, there occur two intraunit excitations, to (a) state T$_3$ which is 1e-1h, LUMO $\to$ LUMO+1 (and HOMO-1 $\to$ HOMO) within the
TIPS-P unit and (b) 2e-2h state T$_4$, where the second excitation is a spin singlet transition across the HOMO - LUMO gap within the other unit
(TIPS-T in PTn and TIPS-P in BPn). Note that the calculated T$_1 \to$ T$_3$ transitions in Fig,~\ref{triplet}(a)-(b) occur at nearly the same wavelengths and with approximately
the same intensities in PTn and BPn, as is expected from Fig.~\ref{triplet}(c). 
The 2e-2h T$_4$ is close in energy to the 1e-1h T$_3$ which is yet again a correlation effect as seen in the case of singlets (see Fig.~\ref{singletESA}). 
The relative energies of the T$_1 \to$ T$_4$ transitions,  
however, are very different, occurring at longer wavelength (lower energy) than the T$_1 \to$ T$_3$ transition in BPn, and at shorter 
wavelength (higher energy) in PTn. This difference is due to the larger HOMO-LUMO gap in the tetracene component of PTn.
\vskip 0.5pc
\begin{figure*}[t]
\includegraphics[width=7.5in]{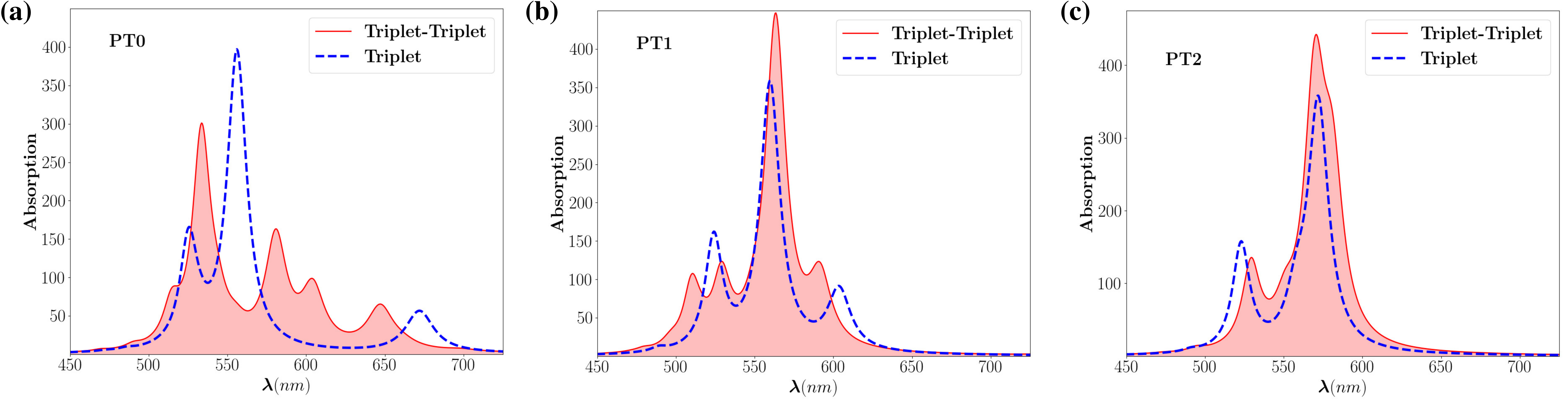}
\caption{Calculated triplet-triplet (solid red) and triplet (broken blue) ESA for (a) PT0, (b) PT1 and (c) PT2 in the visible region.
In PT0, the difference is clear with multiple absorptions taking place in the triplet-triplet subspace. Upon the inclusion of phenyl rings in between the acene monomers
, the distinction is less apparent.
}
\label{TTPA1}
\end{figure*}

\noindent \underbar {\it (c) The Triplet-Triplet eigenstate and ESA.}
The lowest triplet excitons in TIPS-P and TIPS-T monomers occur at 0.89 eV and 1.06 eV, respectively. Naively, the lowest $^1$(TT)$_1$ 
in PT0 and PT1 can occur as a double excitation within the TIPS-P unit of the dimer molecule (note that this possibility does not arise in BPn, where 
the two triplet excitations on different TIPS-P monomers will be of lower energy due to smaller confinement).
From our calculations we find that even in the heterodimer,
the $^1$(TT)$_1$ eigenstate is nearly 80\% 2e-2h (HOMO $\rightarrow$ LUMO)$_P$ $\times$ (HOMO $\rightarrow$ LUMO)$_T$, where the subscripts P and T refer to
TIPS-P and TIPS-T, respectively, 
with weak additional contributions from higher energy 2e-2h configurations 
and even weaker contributions from CT configurations (see reference \onlinecite{sm} for complete $^1$(TT)$_1$ wavefunctions). 
For $U=6.7$ eV and $\kappa=1$, which reproduces the
ground state absorption spectrum quantitatively (see Fig.~\ref{PTn}) we find the $^1$(TT)$_1$ state is above S$_1$ by about 0.3 eV (in contrast to S$^a_1$, the lowest double excitation
within the TIPS-P monomer at $\sim 0.62$ eV above S$_1$, see Fig.~\ref{singletESA}). 
For stronger correlations 
$U=7.7$ eV, $\kappa=1.3$, the calculated $^1$(TT)$_1$ is nearly degenerate with S$_1$. 
Experimentally, observation of delayed fluorescence \cite{Sanders16c}
places the $^1$(TT)$_1$ slightly below the S$_1$.
Our calculated ESAs are therefore for $U=7.7$ eV, $\kappa=1.3$.

There are three fundamental questions we have addressed in the context of $^1$(TT)$_1$  ESA. 
(i) To what
extent is the spin-entanglement in $^1$(TT)$_1$ affected by the nondegeneracies of the triplet excitations within TIPS-P and TIPS-T? Experimentally, this can be revealed from comparisons of ESAs
of PTn and BPn. (ii) What are the natures of the final
states of ESAs, and what do their wavefunctions reveal about correlation effects? (iii) Finally, to  what extent does the $^1$(TT)$_1$ ESA resemble ESA from the free triplet T$_1$ as the
number of phenyl linkers is increased? Experiments indicate decreasing coupling \cite{Sanders15a,Tayebjee17a} between the two triplets with increasing n. 
Our previous calculations for BPn could not be performed for n $>$ 1 because of the large size of the TIPS-pentacene monomer \cite{Khan17b}.

In Figs.~\ref{TTPA} (a) and (b) we compare the calculated ESA spectra of $^1$(TT)$_1$ in BP0 and PT0, respectively.
In Figs.~\ref{TTPA}(c) and (d) we
have shown the dominant components to the final states of the absorptions. 
As seen in Fig.~\ref{TTPA}(c), ESAs from  $^1$(TT)$_1$ in BP0 are of two different kinds, intermonomer CT to $^1$(TT)$_2$ at the lowest energy, and intramonomer 
LUMO$\rightarrow$LUMO+1 transitions at higher energies. 
The intramonomer absorptions
again have two different origins, an intense absorption from the strong 2e-2h component of $^1$(TT)$_1$ to $^1$(TT)$_3$ and a much weaker absorption from the CT contribution
of  $^1$(TT)$_1$ to $^1$(TT)$_4$. 
Comparing against Fig.~\ref{triplet} we see that the
CT transitions to T$_2$ and to $^1$(TT)$_2$ occur at wavelengths that are close, but the intensity is significantly larger for absorption from
$^1$(TT)$_1$. This large difference in intensities has been observed
experimentally \cite{Sanders15a}.

Fig.~\ref{TTPA}(d) explains the origins of the many more $^1$(TT)$_1$ absorptions in PT0 than in BP0. Lifting of the
degeneracies of the HOMO $\to$ LUMO transitions splits several of the transition of Fig.~\ref{TTPA}(c) into two 
(for e.g., the CT transition to $^1$(TT)$_2$ in Fig.~\ref{TTPA}(c) occurs now as two distinct CT transitions to nondegenerate
final states $^1$(TT)$_2$ and $^1$(TT)$_4$); the same is true for the intramonomer transitions). As a consequence of this lifting of degeneracy, the strengths of the individual
transitions in PT0 are considerably weaker. As might be expected from the physical origins of the transitions, all $^1$(TT)$_1$ ESAs, in both BP0 and PT0,    
are polarized predominantly along the long axis of the molecule. 

The consequences of strong correlations can be seen by comparing the relative energies of the intramonomer transitions T$_1$ $\to$ T$_3$ in the triplet ESA of PTn and BPn
in Fig.~\ref{triplet} versus the corresponding transitions to $^1$(TT)$_3$ and $^1$(TT)$_6$ in Fig.~\ref{TTPA}. While the former pair occur virtually at identical energies due to isolated excitations from TIPS-P, the transition to
$^1$(TT)$_6$ in PT0 is observably blushifted relative to the transition to $^1$(TT)$_3$. 
This is because even though both these excitations 
involve only the TIPS-P component, interactions with the neighboring triplet exciton localized on the higher energy TIPS-T unit in PT0 is responsible for the increase in the wavelength of the transition. 
\begin{figure}
\includegraphics[width=3.5in]{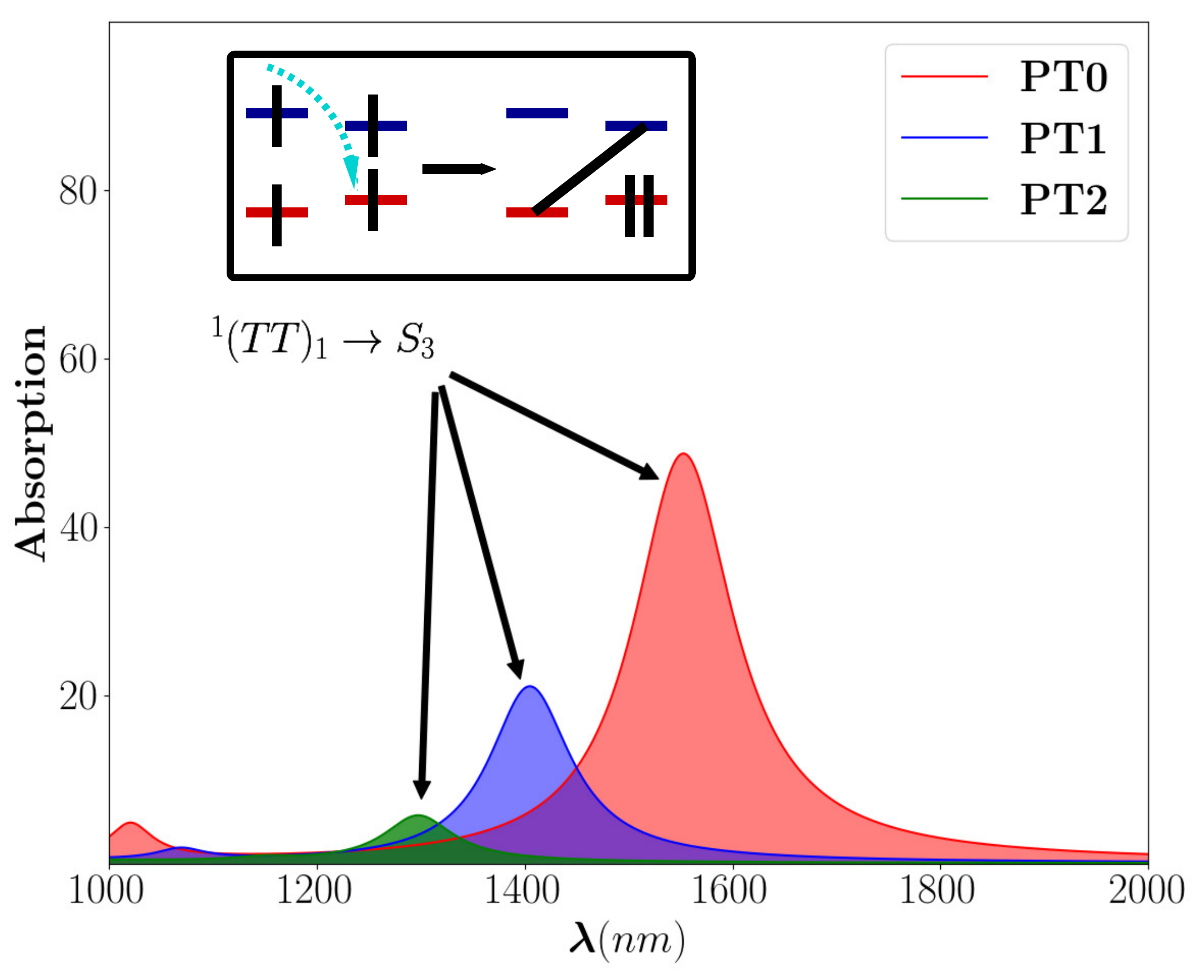}
\caption{Calculated triplet-triplet ESA for PT0 (red), PT1 (blue) and PT2 (green) in the SWIR region ($U=7.7$ eV, $\kappa=1.3$).
Weak optical signals in the infra-red region is reminiscent of the $^1$(TT)$_1$ ESA in bipentacenes, that is absent in the triplet ESA.
(Inset) The final state of the long wavelength absorption is the charge-transfer state S$_3$ which is at a higher energy than the covalent $^1$(TT)$_1$.}
\label{TTPA2}
\end{figure} 
\vskip 0.5pc
\noindent \underbar {\it (d) Entanglement: free triplets versus triplet-triplet.} 
In Figs.~\ref{TTPA1}(a)-(c) we have shown calculated $^1$(TT)$_1$ and T$_1$ ESAs in the visible region superimposed on one another, for PT0, PT1 and PT2,
respectively. The ESA spectra
begin to resemble one another as the number of phenyl spacer groups is increased 
and the two triplets in $^1$(TT)$_1$ become predominantly localized on the monomer units. The overlapping spectra of $^1$(TT)$_1$ and T$_1$ for n = 2
indicate weaker entanglement and confinement of triplets with increasing n.

Finally, we predict an additional absorption from the $^1$(TT)$_1$ at much longer wavelengths,
as shown in Fig.~\ref{TTPA2}, that is completely absent in the free triplet ESA. Fig.~\ref{TTPA2} also indicates the
origin of the absorption: the final states here are the CT states S$_3$ of Fig.~3. While the very occurrence of this absorption is a consequence of the
entanglement in the $^1$(TT)$_1$, once again the decreasing intensity of this absorption indicates decreasing entanglement with increasing n.
\begin{figure*}[t]
\centering
\includegraphics[width=7.5in]{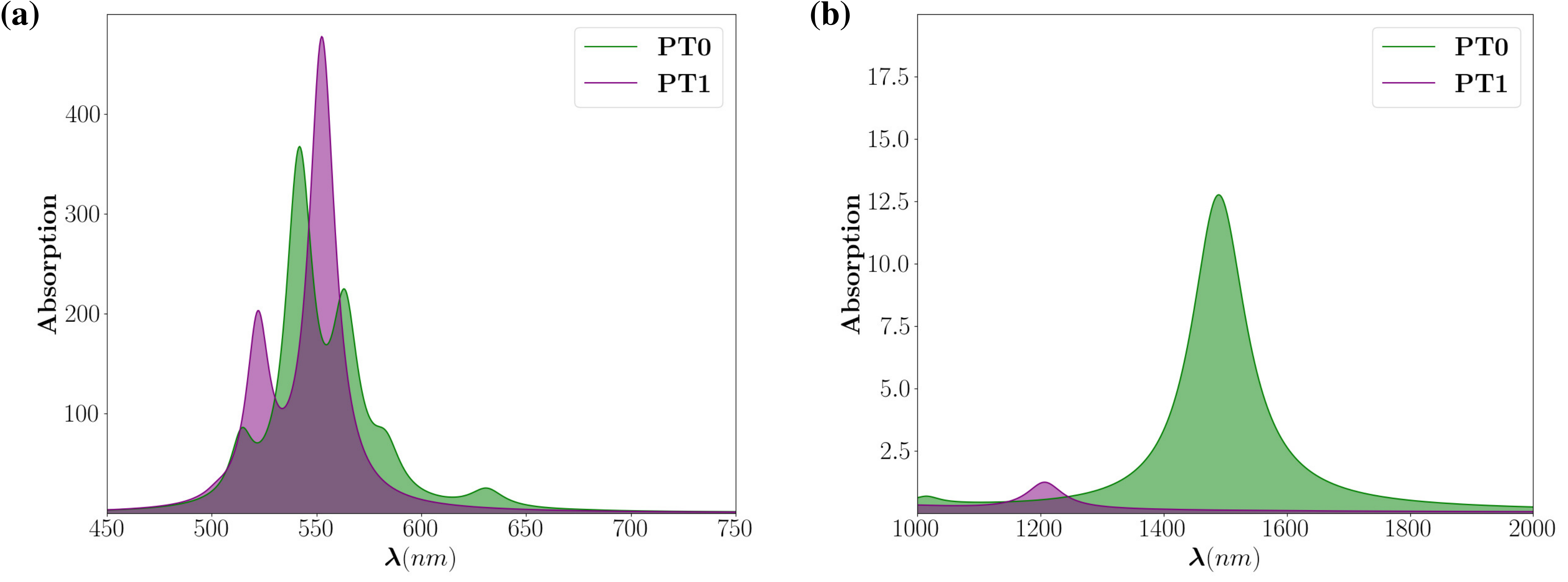}
\caption{Calculated triplet-triplet ESA of PT0 (green) and PT1 (purple) in (a) visible and (b) SWIR.
Effect of rotational twists is explored here with a large dihedral angle between the connecting subunits 
($\theta=60^\circ$ or $t^{inter}_{ij}=-1.1$ eV). 
ESA in the SWIR is dramatically reduced in PTn due to a reduction in the strength of the intermolecular coupling.
Coulomb parameters are chosen to be the following : $U=7.7$ eV and $\kappa=1.3$.
}
\label{TTESA}
\end{figure*}

\noindent \underbar {\it (e) Rotational twists and entanglement}. So far 
we have considered only planar molecular geometries in order to understand the overall trend in the extent the triplet-triplet
entanglement with increase in the number of spacers. Steric hindrance between the phenyl units and the acene monomers implies non-planar geometries
in the real molecules, which in turn implies that $t^{inter}_{ij}$ is smaller than that used in our calculations. 
We have performed calculations of $^1$(TT)$_1$ wavefunctions and ESAs from this state in PT0 and PT1 with $t^{inter}_{ij}(\theta)$ =$-$2.2 cos$\theta$ eV, where 
$t^{inter}_{ij}(\theta)$ is the parametrized hopping integral \cite{Ramasesha91a} for dihedral angle $\theta$. The fundamental $^1$(TT)$_1$ changes very little
from that in Fig.~S8 of reference \onlinecite{sm} in both cases. In Fig.~\ref{TTESA} we show the ESA
spectra corresponding to both PT0 and PT1 for $\theta=60^o$.  
The sharp reduction in the intensities of the excited state absorptions in the SWIR (see Fig.~\ref{TTESA}(b)) in both the dimers is a signature of reduced triplet
entanglement in the real materials. The difference between the triplet and $^1$(TT)$_1$ ESA spectra is vanishingly small now (compare spectra in Fig.~\ref{TTESA}(a) against the triplet ESAs 
in Fig.~\ref{TTPA1}(a)-(b), repsectively).
The decrease in the entanglement is more pronounced in PT1 where a rotation of the phenyl linker leads to decrease in hopping integrals between the phenyl ring and both monomer units.
Only absorptions in the visible are seen in this case, corresponding to intramonomer transitions from the 2e-2h component of $^1$(TT)$_1$. 

\noindent \underbar{\it (f) Polarization dependence of transient absorptions.}
Based on our calculations, aside from S$_1$ $\to$ S$_1^a$ and T$_1$ $\to$ T$_4$, all other TAs are predominantly polarized 
along the long molecular axis of the heterodimer. Hence, polarized TA measurements might be useful in identifying the absorptions in $^1$(TT)$_1$ and S$_1$.
This will have important consequences for interpretations of experimental measurements.
In BPn however, S$^a_1$ is a superposition of 2e-2h and 1e-1h CT excitations. Since, the transition dipole moment due to this 1e-1h component is polarized along the long axis of the molecule
unlike the 2e-2h, the polarization of S$_1$ $\to$ S$_1^a$ is less well defined in BPn than in PTn. Thus with increasing asymmetry, the S$_1$ $\to$ S$_1^a$ 
transition becomes more polarized, because of a decreasing contribution by CT components to the wavefunction.

\section{Conclusion}

By performing full many-body calculations of excited states in a heteroacene dimer we arrive at the following conclusions.

(i) The lowest spin-singlet (S$_1$, S$_2$) and triplet (T$_1$) excitons reside on the individual acene components of the 
heterodimer PTn, with the phenyl linkers playing a negligible role. Similarly, the two triplet excitons of the triplet-triplet
occupy the two acene components. 

(ii) Existing experiments distinguish TAs from the singlet and the triplet-triplet from their lifetimes. Our work indicates that the
singlet will exhibit TA in the long wavelength region that is distinguishably beyond the maximum wavelength where the triplet-triplet ESA occurs.
Thus the issue of spin-entanglement in acene dimers can be studied without complications arising from singlet TA.

(iii) In the absence of rotational twists, the entanglement between the two triplets in $^1$(TT)$_1$ is very strong in PTn with n = 0 and 1. Not only are the TAs from T$_1$ and
$^1$(TT)$_1$ very different in the visible region, the $^1$(TT)$_1$ is predicted to have additional TA in the SWIR. With further increase in n though,
the entanglement is weak. 
With rotational twists between units, the entanglement is less strong, particularly in PT1.

Finally, two important questions emerge from our theoretical work. First, whether the quintet nature of the triplet-triplet in n = 3 dimers \cite{Tayebjee17a}
can be understood theoretically. It has been argued that the separation to free triplets can occur via such quintet states, and the question is clearly
of fundamental and technological interest. Second, what is the nature of the triplet-triplet entanglement in heterodimers of longer acenes, where the intramonomer double 
excitation is at even lower energy, and there can be significant configuration mixing between intra- and intermonomer 2e-2h states? These and other
intriguing topics are currently under investigation. 
\vskip 0.5pc
\section{Acknowledgment}

This work was supported by NSF grant CHE-1764152. 
We would like to thank Arizona TRIF-Photonics for the continuing support.


\begin{thebibliography}{36}%
\makeatletter
\providecommand \@ifxundefined [1]{%
 \@ifx{#1\undefined}
}%
\providecommand \@ifnum [1]{%
 \ifnum #1\expandafter \@firstoftwo
 \else \expandafter \@secondoftwo
 \fi
}%
\providecommand \@ifx [1]{%
 \ifx #1\expandafter \@firstoftwo
 \else \expandafter \@secondoftwo
 \fi
}%
\providecommand \natexlab [1]{#1}%
\providecommand \enquote  [1]{``#1''}%
\providecommand \bibnamefont  [1]{#1}%
\providecommand \bibfnamefont [1]{#1}%
\providecommand \citenamefont [1]{#1}%
\providecommand \href@noop [0]{\@secondoftwo}%
\providecommand \href [0]{\begingroup \@sanitize@url \@href}%
\providecommand \@href[1]{\@@startlink{#1}\@@href}%
\providecommand \@@href[1]{\endgroup#1\@@endlink}%
\providecommand \@sanitize@url [0]{\catcode `\\12\catcode `\$12\catcode
  `\&12\catcode `\#12\catcode `\^12\catcode `\_12\catcode `\%12\relax}%
\providecommand \@@startlink[1]{}%
\providecommand \@@endlink[0]{}%
\providecommand \url  [0]{\begingroup\@sanitize@url \@url }%
\providecommand \@url [1]{\endgroup\@href {#1}{\urlprefix }}%
\providecommand \urlprefix  [0]{URL }%
\providecommand \Eprint [0]{\href }%
\providecommand \doibase [0]{http://dx.doi.org/}%
\providecommand \selectlanguage [0]{\@gobble}%
\providecommand \bibinfo  [0]{\@secondoftwo}%
\providecommand \bibfield  [0]{\@secondoftwo}%
\providecommand \translation [1]{[#1]}%
\providecommand \BibitemOpen [0]{}%
\providecommand \bibitemStop [0]{}%
\providecommand \bibitemNoStop [0]{.\EOS\space}%
\providecommand \EOS [0]{\spacefactor3000\relax}%
\providecommand \BibitemShut  [1]{\csname bibitem#1\endcsname}%
\let\auto@bib@innerbib\@empty
\bibitem [{\citenamefont {Salem}(1966)}]{Salem66a}%
  \BibitemOpen
  \bibfield  {author} {\bibinfo {author} {\bibfnamefont {L.}~\bibnamefont
  {Salem}},\ }\href@noop {} {\emph {\bibinfo {title} {Molecular Orbital Theory
  of Conjugated Systems}}}\ (\bibinfo  {publisher} {Benjamin},\ \bibinfo
  {address} {New York, USA},\ \bibinfo {year} {1966})\BibitemShut {NoStop}%
\bibitem [{\citenamefont {Hudson}\ and\ \citenamefont
  {Kohler}(1973)}]{Kohler73a}%
  \BibitemOpen
  \bibfield  {author} {\bibinfo {author} {\bibfnamefont {B.~S.}\ \bibnamefont
  {Hudson}}\ and\ \bibinfo {author} {\bibfnamefont {B.~E.}\ \bibnamefont
  {Kohler}},\ }\href@noop {} {\bibfield  {journal} {\bibinfo  {journal} {J.\
  Chem.\ Phys.}\ }\textbf {\bibinfo {volume} {59}},\ \bibinfo {pages} {4984}
  (\bibinfo {year} {1973})}\BibitemShut {NoStop}%
\bibitem [{\citenamefont {Hudson}\ \emph {et~al.}(1982)\citenamefont {Hudson},
  \citenamefont {Kohler},\ and\ \citenamefont {Schulten}}]{Hudson82a}%
  \BibitemOpen
  \bibfield  {author} {\bibinfo {author} {\bibfnamefont {B.~S.}\ \bibnamefont
  {Hudson}}, \bibinfo {author} {\bibfnamefont {B.~E.}\ \bibnamefont {Kohler}},
  \ and\ \bibinfo {author} {\bibfnamefont {K.}~\bibnamefont {Schulten}},\
  }\href@noop {} {\bibfield  {journal} {\bibinfo  {journal} {Excited States}\
  }\textbf {\bibinfo {volume} {6}},\ \bibinfo {pages} {1} (\bibinfo {year}
  {1982})}\BibitemShut {NoStop}%
\bibitem [{\citenamefont {Schulten}\ \emph {et~al.}(1976)\citenamefont
  {Schulten}, \citenamefont {Ohmine},\ and\ \citenamefont
  {Karplus}}]{Schulten76a}%
  \BibitemOpen
  \bibfield  {author} {\bibinfo {author} {\bibfnamefont {K.}~\bibnamefont
  {Schulten}}, \bibinfo {author} {\bibfnamefont {I.}~\bibnamefont {Ohmine}}, \
  and\ \bibinfo {author} {\bibfnamefont {M.}~\bibnamefont {Karplus}},\
  }\href@noop {} {\bibfield  {journal} {\bibinfo  {journal} {J.\ Chem.\ Phys.}\
  }\textbf {\bibinfo {volume} {64}},\ \bibinfo {pages} {4422} (\bibinfo {year}
  {1976})}\BibitemShut {NoStop}%
\bibitem [{\citenamefont {Ramasesha}\ and\ \citenamefont
  {Soos}(1984{\natexlab{a}})}]{Ramasesha84b}%
  \BibitemOpen
  \bibfield  {author} {\bibinfo {author} {\bibfnamefont {S.}~\bibnamefont
  {Ramasesha}}\ and\ \bibinfo {author} {\bibfnamefont {Z.~G.}\ \bibnamefont
  {Soos}},\ }\href@noop {} {\bibfield  {journal} {\bibinfo  {journal} {Int. J.
  Quant. Chem.}\ }\textbf {\bibinfo {volume} {25}},\ \bibinfo {pages} {1003}
  (\bibinfo {year} {1984}{\natexlab{a}})}\BibitemShut {NoStop}%
\bibitem [{\citenamefont {Ramasesha}\ and\ \citenamefont
  {Soos}(1984{\natexlab{b}})}]{Ramasesha84c}%
  \BibitemOpen
  \bibfield  {author} {\bibinfo {author} {\bibfnamefont {S.}~\bibnamefont
  {Ramasesha}}\ and\ \bibinfo {author} {\bibfnamefont {Z.~G.}\ \bibnamefont
  {Soos}},\ }\href@noop {} {\bibfield  {journal} {\bibinfo  {journal} {J.\
  Chem.\ Phys.}\ }\textbf {\bibinfo {volume} {80}},\ \bibinfo {pages} {3278}
  (\bibinfo {year} {1984}{\natexlab{b}})}\BibitemShut {NoStop}%
\bibitem [{\citenamefont {Aryanpour}\ \emph {et~al.}(2014)\citenamefont
  {Aryanpour}, \citenamefont {Roberts}, \citenamefont {Sandhu}, \citenamefont
  {Rathore}, \citenamefont {Shukla},\ and\ \citenamefont
  {Mazumdar}}]{Aryanpour14a}%
  \BibitemOpen
  \bibfield  {author} {\bibinfo {author} {\bibfnamefont {K.}~\bibnamefont
  {Aryanpour}}, \bibinfo {author} {\bibfnamefont {A.}~\bibnamefont {Roberts}},
  \bibinfo {author} {\bibfnamefont {A.}~\bibnamefont {Sandhu}}, \bibinfo
  {author} {\bibfnamefont {R.}~\bibnamefont {Rathore}}, \bibinfo {author}
  {\bibfnamefont {A.}~\bibnamefont {Shukla}}, \ and\ \bibinfo {author}
  {\bibfnamefont {S.}~\bibnamefont {Mazumdar}},\ }\href@noop {} {\bibfield
  {journal} {\bibinfo  {journal} {J. Phys. Chem. C}\ }\textbf {\bibinfo
  {volume} {118}},\ \bibinfo {pages} {3331} (\bibinfo {year} {2014})},\ \Eprint
  {http://arxiv.org/abs/http://pubs.acs.org/doi/pdf/10.1021/jp410793r}
  {http://pubs.acs.org/doi/pdf/10.1021/jp410793r} \BibitemShut {NoStop}%
\bibitem [{\citenamefont {Goli}\ \emph {et~al.}(2016)\citenamefont {Goli},
  \citenamefont {Prodhan}, \citenamefont {Mazumdar},\ and\ \citenamefont
  {Ramasesha}}]{Goli16a}%
  \BibitemOpen
  \bibfield  {author} {\bibinfo {author} {\bibfnamefont {V.~M. L. D.~P.}\
  \bibnamefont {Goli}}, \bibinfo {author} {\bibfnamefont {S.}~\bibnamefont
  {Prodhan}}, \bibinfo {author} {\bibfnamefont {S.}~\bibnamefont {Mazumdar}}, \
  and\ \bibinfo {author} {\bibfnamefont {S.}~\bibnamefont {Ramasesha}},\
  }\href@noop {} {\bibfield  {journal} {\bibinfo  {journal} {Phys.\ Rev.\ B}\
  }\textbf {\bibinfo {volume} {94}},\ \bibinfo {pages} {035139} (\bibinfo
  {year} {2016})}\BibitemShut {NoStop}%
\bibitem [{\citenamefont {Smith}\ and\ \citenamefont {Michl}(2013)}]{Smith13a}%
  \BibitemOpen
  \bibfield  {author} {\bibinfo {author} {\bibfnamefont {M.~B.}\ \bibnamefont
  {Smith}}\ and\ \bibinfo {author} {\bibfnamefont {J.}~\bibnamefont {Michl}},\
  }\href@noop {} {\bibfield  {journal} {\bibinfo  {journal} {Annu.\ Rev.\
  Phys.\ Chem.}\ }\textbf {\bibinfo {volume} {64}},\ \bibinfo {pages} {361}
  (\bibinfo {year} {2013})}\BibitemShut {NoStop}%
\bibitem [{\citenamefont {Yong}\ \emph {et~al.}(2017)\citenamefont {Yong},
  \citenamefont {Musser}, \citenamefont {Bayliss}, \citenamefont {Lukman},
  \citenamefont {Tamura}, \citenamefont {Bubnova}, \citenamefont {Hallani},
  \citenamefont {Meneau}, \citenamefont {Resel}, \citenamefont {Maruyama},
  \citenamefont {Hotta}, \citenamefont {Herz}, \citenamefont {Beljonne},
  \citenamefont {Anthony}, \citenamefont {Clark},\ and\ \citenamefont
  {Sirringhaus}}]{Yong17a}%
  \BibitemOpen
  \bibfield  {author} {\bibinfo {author} {\bibfnamefont {C.~K.}\ \bibnamefont
  {Yong}}, \bibinfo {author} {\bibfnamefont {A.~J.}\ \bibnamefont {Musser}},
  \bibinfo {author} {\bibfnamefont {S.~L.}\ \bibnamefont {Bayliss}}, \bibinfo
  {author} {\bibfnamefont {S.}~\bibnamefont {Lukman}}, \bibinfo {author}
  {\bibfnamefont {H.}~\bibnamefont {Tamura}}, \bibinfo {author} {\bibfnamefont
  {O.}~\bibnamefont {Bubnova}}, \bibinfo {author} {\bibfnamefont {R.~K.}\
  \bibnamefont {Hallani}}, \bibinfo {author} {\bibfnamefont {A.}~\bibnamefont
  {Meneau}}, \bibinfo {author} {\bibfnamefont {R.}~\bibnamefont {Resel}},
  \bibinfo {author} {\bibfnamefont {M.}~\bibnamefont {Maruyama}}, \bibinfo
  {author} {\bibfnamefont {S.}~\bibnamefont {Hotta}}, \bibinfo {author}
  {\bibfnamefont {L.~M.}\ \bibnamefont {Herz}}, \bibinfo {author}
  {\bibfnamefont {D.}~\bibnamefont {Beljonne}}, \bibinfo {author}
  {\bibfnamefont {J.~E.}\ \bibnamefont {Anthony}}, \bibinfo {author}
  {\bibfnamefont {J.}~\bibnamefont {Clark}}, \ and\ \bibinfo {author}
  {\bibfnamefont {H.}~\bibnamefont {Sirringhaus}},\ }\href@noop {} {\bibfield
  {journal} {\bibinfo  {journal} {Nat.\ Commun.}\ }\textbf {\bibinfo {volume}
  {8}} (\bibinfo {year} {2017})}\BibitemShut {NoStop}%
\bibitem [{\citenamefont {Stern}\ \emph {et~al.}(2017)\citenamefont {Stern},
  \citenamefont {Cheminal}, \citenamefont {Yost}, \citenamefont {Broch},
  \citenamefont {Bayliss}, \citenamefont {Chen}, \citenamefont {Tabachnyk},
  \citenamefont {Thorley}, \citenamefont {Greenham}, \citenamefont {Hodgkiss},
  \citenamefont {Anthony}, \citenamefont {Head-Gordon}, \citenamefont {Musser},
  \citenamefont {Rao},\ and\ \citenamefont {Friend}}]{Stern17a}%
  \BibitemOpen
  \bibfield  {author} {\bibinfo {author} {\bibfnamefont {H.~L.}\ \bibnamefont
  {Stern}}, \bibinfo {author} {\bibfnamefont {A.}~\bibnamefont {Cheminal}},
  \bibinfo {author} {\bibfnamefont {S.~R.}\ \bibnamefont {Yost}}, \bibinfo
  {author} {\bibfnamefont {K.}~\bibnamefont {Broch}}, \bibinfo {author}
  {\bibfnamefont {S.~L.}\ \bibnamefont {Bayliss}}, \bibinfo {author}
  {\bibfnamefont {K.}~\bibnamefont {Chen}}, \bibinfo {author} {\bibfnamefont
  {M.}~\bibnamefont {Tabachnyk}}, \bibinfo {author} {\bibfnamefont
  {K.}~\bibnamefont {Thorley}}, \bibinfo {author} {\bibfnamefont
  {N.}~\bibnamefont {Greenham}}, \bibinfo {author} {\bibfnamefont {J.~M.}\
  \bibnamefont {Hodgkiss}}, \bibinfo {author} {\bibfnamefont {J.}~\bibnamefont
  {Anthony}}, \bibinfo {author} {\bibfnamefont {M.}~\bibnamefont
  {Head-Gordon}}, \bibinfo {author} {\bibfnamefont {A.~J.}\ \bibnamefont
  {Musser}}, \bibinfo {author} {\bibfnamefont {A.}~\bibnamefont {Rao}}, \ and\
  \bibinfo {author} {\bibfnamefont {R.~H.}\ \bibnamefont {Friend}},\ }\href
  {\doibase 10.1038/nchem.2856} {\bibfield  {journal} {\bibinfo  {journal}
  {Nat. Chem.}\ }\textbf {\bibinfo {volume} {9}},\ \bibinfo {pages} {1205}
  (\bibinfo {year} {2017})}\BibitemShut {NoStop}%
\bibitem [{\citenamefont {Weiss}\ \emph {et~al.}(2016)\citenamefont {Weiss},
  \citenamefont {Bayliss}, \citenamefont {Kraffert}, \citenamefont {Thorley},
  \citenamefont {Anthony}, \citenamefont {Bittl}, \citenamefont {Friend},
  \citenamefont {Rao}, \citenamefont {Greenham},\ and\ \citenamefont
  {Behrends}}]{Weiss16a}%
  \BibitemOpen
  \bibfield  {author} {\bibinfo {author} {\bibfnamefont {L.~R.}\ \bibnamefont
  {Weiss}}, \bibinfo {author} {\bibfnamefont {S.~L.}\ \bibnamefont {Bayliss}},
  \bibinfo {author} {\bibfnamefont {F.}~\bibnamefont {Kraffert}}, \bibinfo
  {author} {\bibfnamefont {K.~J.}\ \bibnamefont {Thorley}}, \bibinfo {author}
  {\bibfnamefont {J.~E.}\ \bibnamefont {Anthony}}, \bibinfo {author}
  {\bibfnamefont {R.}~\bibnamefont {Bittl}}, \bibinfo {author} {\bibfnamefont
  {R.~H.}\ \bibnamefont {Friend}}, \bibinfo {author} {\bibfnamefont
  {A.}~\bibnamefont {Rao}}, \bibinfo {author} {\bibfnamefont {N.~C.}\
  \bibnamefont {Greenham}}, \ and\ \bibinfo {author} {\bibfnamefont
  {J.}~\bibnamefont {Behrends}},\ }\href@noop {} {\bibfield  {journal}
  {\bibinfo  {journal} {Nat. Phys.}\ }\textbf {\bibinfo {volume} {13}},\
  \bibinfo {pages} {176} (\bibinfo {year} {2016})}\BibitemShut {NoStop}%
\bibitem [{\citenamefont {Tayebjee}\ \emph {et~al.}(2017)\citenamefont
  {Tayebjee}, \citenamefont {Sanders}, \citenamefont {Kumaraswamy},
  \citenamefont {Campos}, \citenamefont {Sfeir},\ and\ \citenamefont
  {McCamey}}]{Tayebjee17a}%
  \BibitemOpen
  \bibfield  {author} {\bibinfo {author} {\bibfnamefont {M.~J.~Y.}\
  \bibnamefont {Tayebjee}}, \bibinfo {author} {\bibfnamefont {S.~N.}\
  \bibnamefont {Sanders}}, \bibinfo {author} {\bibfnamefont {E.}~\bibnamefont
  {Kumaraswamy}}, \bibinfo {author} {\bibfnamefont {L.~M.}\ \bibnamefont
  {Campos}}, \bibinfo {author} {\bibfnamefont {M.~Y.}\ \bibnamefont {Sfeir}}, \
  and\ \bibinfo {author} {\bibfnamefont {D.~R.}\ \bibnamefont {McCamey}},\
  }\href@noop {} {\bibfield  {journal} {\bibinfo  {journal} {Nat. Phys.}\
  }\textbf {\bibinfo {volume} {13}},\ \bibinfo {pages} {182} (\bibinfo {year}
  {2017})}\BibitemShut {NoStop}%
\bibitem [{\citenamefont {Basel}\ \emph {et~al.}(2017)\citenamefont {Basel},
  \citenamefont {Zirzlmeier}, \citenamefont {Hetzer}, \citenamefont {Phelan},
  \citenamefont {Krzyaniak}, \citenamefont {Reddy}, \citenamefont {Coto},
  \citenamefont {Horwitz}, \citenamefont {Young}, \citenamefont {White},
  \citenamefont {Hampel}, \citenamefont {Clark}, \citenamefont {Thoss},
  \citenamefont {Tykwinski}, \citenamefont {Wasielewski},\ and\ \citenamefont
  {Guldi}}]{Basel17a}%
  \BibitemOpen
  \bibfield  {author} {\bibinfo {author} {\bibfnamefont {B.~S.}\ \bibnamefont
  {Basel}}, \bibinfo {author} {\bibfnamefont {J.}~\bibnamefont {Zirzlmeier}},
  \bibinfo {author} {\bibfnamefont {C.}~\bibnamefont {Hetzer}}, \bibinfo
  {author} {\bibfnamefont {B.~T.}\ \bibnamefont {Phelan}}, \bibinfo {author}
  {\bibfnamefont {M.~D.}\ \bibnamefont {Krzyaniak}}, \bibinfo {author}
  {\bibfnamefont {S.~R.}\ \bibnamefont {Reddy}}, \bibinfo {author}
  {\bibfnamefont {P.~B.}\ \bibnamefont {Coto}}, \bibinfo {author}
  {\bibfnamefont {N.~E.}\ \bibnamefont {Horwitz}}, \bibinfo {author}
  {\bibfnamefont {R.~M.}\ \bibnamefont {Young}}, \bibinfo {author}
  {\bibfnamefont {F.~J.}\ \bibnamefont {White}}, \bibinfo {author}
  {\bibfnamefont {F.}~\bibnamefont {Hampel}}, \bibinfo {author} {\bibfnamefont
  {T.}~\bibnamefont {Clark}}, \bibinfo {author} {\bibfnamefont
  {M.}~\bibnamefont {Thoss}}, \bibinfo {author} {\bibfnamefont {R.~R.}\
  \bibnamefont {Tykwinski}}, \bibinfo {author} {\bibfnamefont {M.~R.}\
  \bibnamefont {Wasielewski}}, \ and\ \bibinfo {author} {\bibfnamefont {D.~M.}\
  \bibnamefont {Guldi}},\ }\href {\doibase 10.1038/ncomms15171} {\bibfield
  {journal} {\bibinfo  {journal} {Nat.\ Commun.}\ }\textbf {\bibinfo {volume}
  {8}} (\bibinfo {year} {2017}),\ 10.1038/ncomms15171}\BibitemShut {NoStop}%
\bibitem [{\citenamefont {Sanders}\ \emph {et~al.}(2016)\citenamefont
  {Sanders}, \citenamefont {Kumarasamy}, \citenamefont {Pun}, \citenamefont
  {Appavoo}, \citenamefont {Steigerwald}, \citenamefont {Campos},\ and\
  \citenamefont {Sfeir}}]{Sanders16c}%
  \BibitemOpen
  \bibfield  {author} {\bibinfo {author} {\bibfnamefont {S.~N.}\ \bibnamefont
  {Sanders}}, \bibinfo {author} {\bibfnamefont {E.}~\bibnamefont {Kumarasamy}},
  \bibinfo {author} {\bibfnamefont {A.~B.}\ \bibnamefont {Pun}}, \bibinfo
  {author} {\bibfnamefont {K.}~\bibnamefont {Appavoo}}, \bibinfo {author}
  {\bibfnamefont {M.~L.}\ \bibnamefont {Steigerwald}}, \bibinfo {author}
  {\bibfnamefont {L.~M.}\ \bibnamefont {Campos}}, \ and\ \bibinfo {author}
  {\bibfnamefont {M.~Y.}\ \bibnamefont {Sfeir}},\ }\href@noop {} {\bibfield
  {journal} {\bibinfo  {journal} {J.\ Am.\ Chem.\ Soc.}\ }\textbf {\bibinfo
  {volume} {138}},\ \bibinfo {pages} {7289} (\bibinfo {year}
  {2016})}\BibitemShut {NoStop}%
\bibitem [{\citenamefont {Chandross}\ \emph {et~al.}(1999)\citenamefont
  {Chandross}, \citenamefont {Shimoi},\ and\ \citenamefont
  {Mazumdar}}]{Chandross99a}%
  \BibitemOpen
  \bibfield  {author} {\bibinfo {author} {\bibfnamefont {M.}~\bibnamefont
  {Chandross}}, \bibinfo {author} {\bibfnamefont {Y.}~\bibnamefont {Shimoi}}, \
  and\ \bibinfo {author} {\bibfnamefont {S.}~\bibnamefont {Mazumdar}},\
  }\href@noop {} {\bibfield  {journal} {\bibinfo  {journal} {Phys.\ Rev.\ B}\
  }\textbf {\bibinfo {volume} {59}},\ \bibinfo {pages} {4822} (\bibinfo {year}
  {1999})}\BibitemShut {NoStop}%
\bibitem [{\citenamefont {Khan}\ and\ \citenamefont
  {Mazumdar}(2017{\natexlab{a}})}]{Khan17b}%
  \BibitemOpen
  \bibfield  {author} {\bibinfo {author} {\bibfnamefont {S.}~\bibnamefont
  {Khan}}\ and\ \bibinfo {author} {\bibfnamefont {S.}~\bibnamefont
  {Mazumdar}},\ }\href {\doibase 10.1021/acs.jpclett.7b01829} {\bibfield
  {journal} {\bibinfo  {journal} {J. Phys. Chem. Lett.}\ }\textbf {\bibinfo
  {volume} {8}},\ \bibinfo {pages} {4468} (\bibinfo {year}
  {2017}{\natexlab{a}})}\BibitemShut {NoStop}%
\bibitem [{\citenamefont {Khan}\ and\ \citenamefont
  {Mazumdar}(2017{\natexlab{b}})}]{Khan17c}%
  \BibitemOpen
  \bibfield  {author} {\bibinfo {author} {\bibfnamefont {S.}~\bibnamefont
  {Khan}}\ and\ \bibinfo {author} {\bibfnamefont {S.}~\bibnamefont
  {Mazumdar}},\ }\href {\doibase 10.1021/acs.jpclett.7b02748.} {\bibfield
  {journal} {\bibinfo  {journal} {J. Phys. Chem. Lett.}\ }\textbf {\bibinfo
  {volume} {8}},\ \bibinfo {pages} {5943} (\bibinfo {year}
  {2017}{\natexlab{b}})}\BibitemShut {NoStop}%
\bibitem [{\citenamefont {Pariser}\ and\ \citenamefont
  {Parr}(1953)}]{Pariser53a}%
  \BibitemOpen
  \bibfield  {author} {\bibinfo {author} {\bibfnamefont {R.}~\bibnamefont
  {Pariser}}\ and\ \bibinfo {author} {\bibfnamefont {R.}~\bibnamefont {Parr}},\
  }\href@noop {} {\bibfield  {journal} {\bibinfo  {journal} {J.\ Chem.\ Phys.}\
  }\textbf {\bibinfo {volume} {21}},\ \bibinfo {pages} {767} (\bibinfo {year}
  {1953})}\BibitemShut {NoStop}%
\bibitem [{\citenamefont {Pople}(1953)}]{Pople53a}%
  \BibitemOpen
  \bibfield  {author} {\bibinfo {author} {\bibfnamefont {J.~A.}\ \bibnamefont
  {Pople}},\ }\href@noop {} {\bibfield  {journal} {\bibinfo  {journal} {Trans.
  Faraday Soc.}\ }\textbf {\bibinfo {volume} {49}},\ \bibinfo {pages} {1375}
  (\bibinfo {year} {1953})}\BibitemShut {NoStop}%
\bibitem [{\citenamefont {Houk}\ \emph {et~al.}(2001)\citenamefont {Houk},
  \citenamefont {Lee},\ and\ \citenamefont {Nendel}}]{Houk01a}%
  \BibitemOpen
  \bibfield  {author} {\bibinfo {author} {\bibfnamefont {K.~N.}\ \bibnamefont
  {Houk}}, \bibinfo {author} {\bibfnamefont {P.~S.}\ \bibnamefont {Lee}}, \
  and\ \bibinfo {author} {\bibfnamefont {M.}~\bibnamefont {Nendel}},\
  }\href@noop {} {\bibfield  {journal} {\bibinfo  {journal} {J.\ Org.\ Chem.}\
  }\textbf {\bibinfo {volume} {66}},\ \bibinfo {pages} {5517} (\bibinfo {year}
  {2001})},\ \Eprint
  {http://arxiv.org/abs/http://pubs.acs.org/doi/pdf/10.1021/jo010391f}
  {http://pubs.acs.org/doi/pdf/10.1021/jo010391f} \BibitemShut {NoStop}%
\bibitem [{\citenamefont {Ducasse}\ \emph {et~al.}(1982)\citenamefont
  {Ducasse}, \citenamefont {Miller},\ and\ \citenamefont {Soos}}]{Ducasse82a}%
  \BibitemOpen
  \bibfield  {author} {\bibinfo {author} {\bibfnamefont {L.~R.}\ \bibnamefont
  {Ducasse}}, \bibinfo {author} {\bibfnamefont {T.~E.}\ \bibnamefont {Miller}},
  \ and\ \bibinfo {author} {\bibfnamefont {Z.~G.}\ \bibnamefont {Soos}},\
  }\href@noop {} {\bibfield  {journal} {\bibinfo  {journal} {J.\ Chem.\ Phys.}\
  }\textbf {\bibinfo {volume} {76}},\ \bibinfo {pages} {4094} (\bibinfo {year}
  {1982})}\BibitemShut {NoStop}%
\bibitem [{\citenamefont {Chandross}\ and\ \citenamefont
  {Mazumdar}(1997)}]{Chandross97a}%
  \BibitemOpen
  \bibfield  {author} {\bibinfo {author} {\bibfnamefont {M.}~\bibnamefont
  {Chandross}}\ and\ \bibinfo {author} {\bibfnamefont {S.}~\bibnamefont
  {Mazumdar}},\ }\href@noop {} {\bibfield  {journal} {\bibinfo  {journal}
  {Phys.\ Rev.\ B}\ }\textbf {\bibinfo {volume} {55}},\ \bibinfo {pages} {1497}
  (\bibinfo {year} {1997})}\BibitemShut {NoStop}%
\bibitem [{sm()}]{sm}%
  \BibitemOpen
  \href@noop {} {}\bibinfo {note} {See Supplemental Material at
  http://link.aps.org/ supplemental/xx.xxxx/ PhysRevB.xxx.xxxxxx for further
  details of calculations.}\BibitemShut {Stop}%
\bibitem [{\citenamefont {Tavan}\ and\ \citenamefont
  {Schulten}(1987)}]{Tavan87a}%
  \BibitemOpen
  \bibfield  {author} {\bibinfo {author} {\bibfnamefont {P.}~\bibnamefont
  {Tavan}}\ and\ \bibinfo {author} {\bibfnamefont {K.}~\bibnamefont
  {Schulten}},\ }\href@noop {} {\bibfield  {journal} {\bibinfo  {journal}
  {Phys.\ Rev.\ B}\ }\textbf {\bibinfo {volume} {36}},\ \bibinfo {pages} {4337}
  (\bibinfo {year} {1987})}\BibitemShut {NoStop}%
\bibitem [{\citenamefont {Platt}\ \emph {et~al.}(2009)\citenamefont {Platt},
  \citenamefont {Day}, \citenamefont {Subramanian}, \citenamefont {Anthony},\
  and\ \citenamefont {Ostroverkhova}}]{Platt09a}%
  \BibitemOpen
  \bibfield  {author} {\bibinfo {author} {\bibfnamefont {A.~D.}\ \bibnamefont
  {Platt}}, \bibinfo {author} {\bibfnamefont {J.}~\bibnamefont {Day}}, \bibinfo
  {author} {\bibfnamefont {S.}~\bibnamefont {Subramanian}}, \bibinfo {author}
  {\bibfnamefont {J.~E.}\ \bibnamefont {Anthony}}, \ and\ \bibinfo {author}
  {\bibfnamefont {O.}~\bibnamefont {Ostroverkhova}},\ }\href@noop {} {\bibfield
   {journal} {\bibinfo  {journal} {J. Phys. Chem. C}\ }\textbf {\bibinfo
  {volume} {113}},\ \bibinfo {pages} {14006} (\bibinfo {year}
  {2009})}\BibitemShut {NoStop}%
\bibitem [{\citenamefont {Ramanan}\ \emph {et~al.}(2012)\citenamefont
  {Ramanan}, \citenamefont {Smeigh}, \citenamefont {Anthony}, \citenamefont
  {Marks},\ and\ \citenamefont {Wasielewski}}]{Ramanan12a}%
  \BibitemOpen
  \bibfield  {author} {\bibinfo {author} {\bibfnamefont {C.}~\bibnamefont
  {Ramanan}}, \bibinfo {author} {\bibfnamefont {A.~L.}\ \bibnamefont {Smeigh}},
  \bibinfo {author} {\bibfnamefont {J.~E.}\ \bibnamefont {Anthony}}, \bibinfo
  {author} {\bibfnamefont {T.~J.}\ \bibnamefont {Marks}}, \ and\ \bibinfo
  {author} {\bibfnamefont {M.~R.}\ \bibnamefont {Wasielewski}},\ }\href@noop {}
  {\bibfield  {journal} {\bibinfo  {journal} {J.\ Am.\ Chem.\ Soc.}\ }\textbf
  {\bibinfo {volume} {134}},\ \bibinfo {pages} {386} (\bibinfo {year}
  {2012})},\ \Eprint
  {http://arxiv.org/abs/http://pubs.acs.org/doi/pdf/10.1021/ja2080482}
  {http://pubs.acs.org/doi/pdf/10.1021/ja2080482} \BibitemShut {NoStop}%
\bibitem [{\citenamefont {Stern}\ \emph {et~al.}(2015)\citenamefont {Stern},
  \citenamefont {Musser}, \citenamefont {Gelinas}, \citenamefont {Parkinson},
  \citenamefont {Herz}, \citenamefont {Bruzek}, \citenamefont {Anthony},
  \citenamefont {Friend},\ and\ \citenamefont {Walker}}]{Stern15a}%
  \BibitemOpen
  \bibfield  {author} {\bibinfo {author} {\bibfnamefont {H.~L.}\ \bibnamefont
  {Stern}}, \bibinfo {author} {\bibfnamefont {A.~J.}\ \bibnamefont {Musser}},
  \bibinfo {author} {\bibfnamefont {S.}~\bibnamefont {Gelinas}}, \bibinfo
  {author} {\bibfnamefont {P.}~\bibnamefont {Parkinson}}, \bibinfo {author}
  {\bibfnamefont {L.~M.}\ \bibnamefont {Herz}}, \bibinfo {author}
  {\bibfnamefont {M.~J.}\ \bibnamefont {Bruzek}}, \bibinfo {author}
  {\bibfnamefont {J.}~\bibnamefont {Anthony}}, \bibinfo {author} {\bibfnamefont
  {R.~H.}\ \bibnamefont {Friend}}, \ and\ \bibinfo {author} {\bibfnamefont
  {B.~J.}\ \bibnamefont {Walker}},\ }\href@noop {} {\bibfield  {journal}
  {\bibinfo  {journal} {Proc. Natl. Acad. Sci.}\ }\textbf {\bibinfo {volume}
  {112}},\ \bibinfo {pages} {7656} (\bibinfo {year} {2015})}\BibitemShut
  {NoStop}%
\bibitem [{\citenamefont {Sony}\ and\ \citenamefont {Shukla}(2007)}]{Sony07a}%
  \BibitemOpen
  \bibfield  {author} {\bibinfo {author} {\bibfnamefont {P.}~\bibnamefont
  {Sony}}\ and\ \bibinfo {author} {\bibfnamefont {A.}~\bibnamefont {Shukla}},\
  }\href@noop {} {\bibfield  {journal} {\bibinfo  {journal} {Phys. Rev. B}\
  }\textbf {\bibinfo {volume} {75}},\ \bibinfo {pages} {155208} (\bibinfo
  {year} {2007})}\BibitemShut {NoStop}%
\bibitem [{\citenamefont {Psiachos}\ and\ \citenamefont
  {Mazumdar}(2009)}]{Psiachos09a}%
  \BibitemOpen
  \bibfield  {author} {\bibinfo {author} {\bibfnamefont {D.}~\bibnamefont
  {Psiachos}}\ and\ \bibinfo {author} {\bibfnamefont {S.}~\bibnamefont
  {Mazumdar}},\ }\href@noop {} {\bibfield  {journal} {\bibinfo  {journal}
  {Phys. Rev. B}\ }\textbf {\bibinfo {volume} {79}},\ \bibinfo {pages} {155106}
  (\bibinfo {year} {2009})}\BibitemShut {NoStop}%
\bibitem [{\citenamefont {Trinh}\ \emph {et~al.}(2017)\citenamefont {Trinh},
  \citenamefont {Pinkard}, \citenamefont {Pun}, \citenamefont {Sanders},
  \citenamefont {Kumarasamy}, \citenamefont {Sfeir}, \citenamefont {Campos},
  \citenamefont {Roy},\ and\ \citenamefont {Zhu}}]{Trinh17a}%
  \BibitemOpen
  \bibfield  {author} {\bibinfo {author} {\bibfnamefont {M.~T.}\ \bibnamefont
  {Trinh}}, \bibinfo {author} {\bibfnamefont {A.}~\bibnamefont {Pinkard}},
  \bibinfo {author} {\bibfnamefont {A.~B.}\ \bibnamefont {Pun}}, \bibinfo
  {author} {\bibfnamefont {S.~N.}\ \bibnamefont {Sanders}}, \bibinfo {author}
  {\bibfnamefont {E.}~\bibnamefont {Kumarasamy}}, \bibinfo {author}
  {\bibfnamefont {M.~Y.}\ \bibnamefont {Sfeir}}, \bibinfo {author}
  {\bibfnamefont {L.~M.}\ \bibnamefont {Campos}}, \bibinfo {author}
  {\bibfnamefont {X.}~\bibnamefont {Roy}}, \ and\ \bibinfo {author}
  {\bibfnamefont {X.-Y.}\ \bibnamefont {Zhu}},\ }\href {\doibase
  10.1126/sciadv.1700241} {\bibfield  {journal} {\bibinfo  {journal} {Science
  Advances}\ }\textbf {\bibinfo {volume} {3}} (\bibinfo {year} {2017}),\
  10.1126/sciadv.1700241},\ \Eprint {http://arxiv.org/abs/e1700241} {e1700241}
  \BibitemShut {NoStop}%
\bibitem [{\citenamefont {Pensack}\ \emph {et~al.}(2016)\citenamefont
  {Pensack}, \citenamefont {Ostroumov}, \citenamefont {Tilley}, \citenamefont
  {Mazza}, \citenamefont {Grieco}, \citenamefont {Thorley}, \citenamefont
  {Asbury}, \citenamefont {Seferos}, \citenamefont {Anthony},\ and\
  \citenamefont {Scholes}}]{Pensack16a}%
  \BibitemOpen
  \bibfield  {author} {\bibinfo {author} {\bibfnamefont {R.~D.}\ \bibnamefont
  {Pensack}}, \bibinfo {author} {\bibfnamefont {E.~E.}\ \bibnamefont
  {Ostroumov}}, \bibinfo {author} {\bibfnamefont {A.~J.}\ \bibnamefont
  {Tilley}}, \bibinfo {author} {\bibfnamefont {S.}~\bibnamefont {Mazza}},
  \bibinfo {author} {\bibfnamefont {C.}~\bibnamefont {Grieco}}, \bibinfo
  {author} {\bibfnamefont {K.~J.}\ \bibnamefont {Thorley}}, \bibinfo {author}
  {\bibfnamefont {J.~B.}\ \bibnamefont {Asbury}}, \bibinfo {author}
  {\bibfnamefont {D.~S.}\ \bibnamefont {Seferos}}, \bibinfo {author}
  {\bibfnamefont {J.~E.}\ \bibnamefont {Anthony}}, \ and\ \bibinfo {author}
  {\bibfnamefont {G.~D.}\ \bibnamefont {Scholes}},\ }\href@noop {} {\bibfield
  {journal} {\bibinfo  {journal} {J. Phys. Chem. Lett.}\ }\textbf {\bibinfo
  {volume} {7}},\ \bibinfo {pages} {2370} (\bibinfo {year} {2016})}\BibitemShut
  {NoStop}%
\bibitem [{\citenamefont {Walker}\ \emph {et~al.}(2013)\citenamefont {Walker},
  \citenamefont {Musser}, \citenamefont {Beljonne},\ and\ \citenamefont
  {Friend}}]{Walker13a}%
  \BibitemOpen
  \bibfield  {author} {\bibinfo {author} {\bibfnamefont {B.~J.}\ \bibnamefont
  {Walker}}, \bibinfo {author} {\bibfnamefont {A.~J.}\ \bibnamefont {Musser}},
  \bibinfo {author} {\bibfnamefont {D.}~\bibnamefont {Beljonne}}, \ and\
  \bibinfo {author} {\bibfnamefont {R.~H.}\ \bibnamefont {Friend}},\
  }\href@noop {} {\bibfield  {journal} {\bibinfo  {journal} {Nat. Chem.}\
  }\textbf {\bibinfo {volume} {5}},\ \bibinfo {pages} {1019} (\bibinfo {year}
  {2013})}\BibitemShut {NoStop}%
\bibitem [{\citenamefont {Grieco}\ \emph {et~al.}(2018)\citenamefont {Grieco},
  \citenamefont {Kennehan}, \citenamefont {Kim}, \citenamefont {Pensack},
  \citenamefont {Brigeman}, \citenamefont {Rimshaw}, \citenamefont {Payne},
  \citenamefont {Anthony}, \citenamefont {Giebink}, \citenamefont {Scholes},\
  and\ \citenamefont {Asbury}}]{Grieco18a}%
  \BibitemOpen
  \bibfield  {author} {\bibinfo {author} {\bibfnamefont {C.}~\bibnamefont
  {Grieco}}, \bibinfo {author} {\bibfnamefont {E.~R.}\ \bibnamefont
  {Kennehan}}, \bibinfo {author} {\bibfnamefont {H.}~\bibnamefont {Kim}},
  \bibinfo {author} {\bibfnamefont {R.~D.}\ \bibnamefont {Pensack}}, \bibinfo
  {author} {\bibfnamefont {A.~N.}\ \bibnamefont {Brigeman}}, \bibinfo {author}
  {\bibfnamefont {A.}~\bibnamefont {Rimshaw}}, \bibinfo {author} {\bibfnamefont
  {M.~M.}\ \bibnamefont {Payne}}, \bibinfo {author} {\bibfnamefont {J.~E.}\
  \bibnamefont {Anthony}}, \bibinfo {author} {\bibfnamefont {N.~C.}\
  \bibnamefont {Giebink}}, \bibinfo {author} {\bibfnamefont {G.~D.}\
  \bibnamefont {Scholes}}, \ and\ \bibinfo {author} {\bibfnamefont {J.~B.}\
  \bibnamefont {Asbury}},\ }\href@noop {} {\bibfield  {journal} {\bibinfo
  {journal} {J. Phys. Chem. C}\ }\textbf {\bibinfo {volume} {122}},\ \bibinfo
  {pages} {2012} (\bibinfo {year} {2018})}\BibitemShut {NoStop}%
\bibitem [{\citenamefont {Sanders}\ \emph {et~al.}(2015)\citenamefont
  {Sanders}, \citenamefont {Kumarasamy}, \citenamefont {Pun}, \citenamefont
  {Trinh}, \citenamefont {Choi}, \citenamefont {Xia}, \citenamefont {Taffet},
  \citenamefont {Low}, \citenamefont {Miller}, \citenamefont {Roy},
  \citenamefont {Zhu}, \citenamefont {Steigerwald}, \citenamefont {Sfeir},\
  and\ \citenamefont {Campos}}]{Sanders15a}%
  \BibitemOpen
  \bibfield  {author} {\bibinfo {author} {\bibfnamefont {S.~N.}\ \bibnamefont
  {Sanders}}, \bibinfo {author} {\bibfnamefont {E.}~\bibnamefont {Kumarasamy}},
  \bibinfo {author} {\bibfnamefont {A.~B.}\ \bibnamefont {Pun}}, \bibinfo
  {author} {\bibfnamefont {M.~T.}\ \bibnamefont {Trinh}}, \bibinfo {author}
  {\bibfnamefont {B.}~\bibnamefont {Choi}}, \bibinfo {author} {\bibfnamefont
  {J.}~\bibnamefont {Xia}}, \bibinfo {author} {\bibfnamefont {E.~J.}\
  \bibnamefont {Taffet}}, \bibinfo {author} {\bibfnamefont {J.~Z.}\
  \bibnamefont {Low}}, \bibinfo {author} {\bibfnamefont {J.~R.}\ \bibnamefont
  {Miller}}, \bibinfo {author} {\bibfnamefont {X.}~\bibnamefont {Roy}},
  \bibinfo {author} {\bibfnamefont {X.-Y.}\ \bibnamefont {Zhu}}, \bibinfo
  {author} {\bibfnamefont {M.~L.}\ \bibnamefont {Steigerwald}}, \bibinfo
  {author} {\bibfnamefont {M.~Y.}\ \bibnamefont {Sfeir}}, \ and\ \bibinfo
  {author} {\bibfnamefont {L.~M.}\ \bibnamefont {Campos}},\ }\href {\doibase
  10.1021/jacs.5b04986} {\bibfield  {journal} {\bibinfo  {journal} {J.\ Am.\
  Chem.\ Soc.}\ }\textbf {\bibinfo {volume} {137}},\ \bibinfo {pages} {8965}
  (\bibinfo {year} {2015})},\ \Eprint
  {http://arxiv.org/abs/http://dx.doi.org/10.1021/jacs.5b04986}
  {http://dx.doi.org/10.1021/jacs.5b04986} \BibitemShut {NoStop}%
\bibitem [{\citenamefont {Ramasesha}\ \emph {et~al.}(1991)\citenamefont
  {Ramasesha}, \citenamefont {Albert},\ and\ \citenamefont
  {Sinha}}]{Ramasesha91a}%
  \BibitemOpen
  \bibfield  {author} {\bibinfo {author} {\bibfnamefont {S.}~\bibnamefont
  {Ramasesha}}, \bibinfo {author} {\bibfnamefont {I.~D.~L.}\ \bibnamefont
  {Albert}}, \ and\ \bibinfo {author} {\bibfnamefont {B.}~\bibnamefont
  {Sinha}},\ }\href@noop {} {\bibfield  {journal} {\bibinfo  {journal} {Mol.
  Phys.}\ }\textbf {\bibinfo {volume} {72}},\ \bibinfo {pages} {537} (\bibinfo
  {year} {1991})}\BibitemShut {NoStop}%
\end{thebibliography}

%

\end{document}